\documentclass[fleqn,11pt]{article}

\usepackage{amsfonts,amsmath,amssymb}
\usepackage{latexsym}
\usepackage{float}
\input preamble.cls

\usepackage[english]{babel}
\usepackage{microtype}
\usepackage[T1,T2A]{fontenc}

\usepackage{microtype}
\usepackage{microtype}
\usepackage{stfloats}

\begin{document}
\selectlanguage{english}
\twocolumn[
\jnumber{issue}{year}

\Title{The Nature of Small-Scale Perturbation Modes in a Nonstationary Self-Gravitating Disk}

\Author{K.T.Mirtadjieva\au{1,2*}, J.M.Ganiev\au{2}, S.N.Nuritdinov\au{2}}
 	     {Ulugh Beg Astronomical Institute, Uzbekistan Academy of Sciences, 33 Astronomicheskaya Street, Tashkent 100052, Uzbekistan$^{1}$, \\
National University of Uzbekistan named after Mirzo Ulugbek, Tashkent, 100174, Uzbekistan$^{2}$}  	   


\Abstract
        {In this work, we investigate the nature of gravitational instability of small-scale perturbation modes during the early nonstationary stages of the evolution of disk-like galaxies. To this end, we study the gravitational instability of small-scale horizontal perturbation modes, such as $(m;N) = (12;20)$, $(14;20)$, $(16;20)$, and $(18;20)$, in the framework of a nonlinear nonstationary anisotropic model of a self-gravitating disk. The corresponding nonstationary analogues of the dispersion equation for these perturbation modes are derived, and their numerical analysis is performed. The results are presented in the form of critical dependences of the initial virial ratio on the disk rotation parameter. The nature of the instability of small-scale perturbation modes in a nonlinear nonstationary self-gravitating disk is determined, and a comparative analysis of their instability increments is carried out.}

\medskip
] 
\email 1 {mkt1959@mail.ru}
\email 1 {karomat@astrin.uz}

\section{Introduction}
The structure of the disk subsystems of galaxies is highly diverse encompassing both large-scale and various small-scale structures. Small-scale structures in galactic disks include open star clusters (OSC), molecular clouds (MC), gas-dust nebulae and other formations. 
The formation processes of these structures and their corresponding primordial protoclouds can be partially explained by the instabilities of specific perturbation modes since one of the fundamental physical mechanisms responsible for the formation of structures in galaxies and other self-gravitating systems is gravitational instability [1–5].
In Refs. [6,7] moderately small-scale modes $(N=10;,m=4,6,8)$ were investigated for “hot” rigidly rotating disks based on the equilibrium model of Bisnovatyi-Kogan and Zel'dovich [8]. Exact characteristic equations for the eigenfrequencies and explicit expressions for the eigenfunctions were obtained. Solutions were found not only for small-scale but also for large-scale perturbations and not only for “nearly cold” systems with nearly circular orbits but also for “hot” systems in which the radial and circular velocities of the particles are of the same order of magnitude.
However these instabilities were investigated only for the strictly equilibrium state of a self-gravitating disk. In contrast the observed structural formations in galaxies develop during the nonlinear nonstationary stage of their evolution. Among the various types of nonstationarity radial pulsation of a self-gravitating disk plays a particularly important role [7,9]. Motivated by this nonlinear pulsating models of self-gravitating disks with isotropic and anisotropic velocity distributions were developed in Refs. [9–13]. Within the framework of these models gravitational instabilities of horizontal and vertical large-scale perturbation modes were investigated in order to clarify the physical origin of large-scale structural formations in galaxies (see, e.g., [10–25]). 

As noted above from the point of view of the smallest-scale perturbation modes the main structural formations in the disk subsystems of galaxies are open star clusters (OSC) and molecular clouds (MC). These structures contribute to the global structure of galactic disks. However the problem of their formation in disk-like galaxies particularly in our Galaxy has not yet been adequately analyzed. This raises a number of important questions can their global distribution be explained by an appropriate formation theory? Under what physical conditions can these objects form in disk-like systems and what are the characteristic timescales of these processes?
To address these questions a separate analysis of the gravitational instability of small-scale perturbation modes against the background of a radially pulsating model of a self-gravitating disk is required. Accordingly in this work we investigate the gravitational instability of horizontal small-scale perturbation modes $(m;N)=(12;20)$, $(14;20)$, $(16;20)$ and $(18;20)$ against the background of a nonlinear pulsating anisotropic model of a self-gravitating disk. The corresponding nonstationary analogues of the dispersion equations (NADE) for these perturbation modes are obtained and their numerical analysis is performed. Critical dependences of the initial virial ratio on the disk rotation parameter are constructed. The physical conditions for the formation of structures associated with the instability of small-scale perturbation modes in a nonlinear pulsating disk are determined. A comparative analysis of the gravitational instability increments of the considered small-scale perturbation modes is also performed for different values of the rotation parameter of the self-gravitating disk.

\section{Basic Relations and Equations}
As noted above one of the main types of nonstationarity is associated primarily with the global radial motions of the system as a whole. Based on this consideration a nonlinear pulsating phase-space disk model was first constructed in Refs. [9,10] in the form

\bearr
    \Psi(r,{{v}_{r}},{{v}_{\bot }},t ) = \frac{\sigma_{0}}{2\pi\Pi \sqrt{1-\Omega^2}} \Bigg[\frac{1-{{\Omega}^{2}}}{{{\Pi}^{2}}} \Bigg(1-\frac{{{r}^{2}}}{{{\Pi}^{2}}}\Bigg)-
\nnn
    -{{({{v}_{r}}-{{v}_{a}})}^{2}}-{{({{v}_{\bot}}-{{v}_{b}})}^{2}} \Bigg]^{-\frac{1}{2}}\cdot \chi(R-r).
\ear

\noindent which represents a nonstationary generalization of the well-known equilibrium isotropic model of Bisnovatyi-Kogan and Zel'dovich [8]. The self-gravitating disk (1) undergoes radial pulsations according to the law $R(t)=R_{0} \Pi (t)$, where
\beq
    \Pi\left(t\right)=\frac{1+\lambda \cos \psi }{1-\lambda ^{2} }  ,     \quad  \quad  \quad     t=\frac{\psi +\lambda \sin \psi }{\left(1-\lambda ^{2} \right)^{3/2} }  ,                       
\eeq
where the amplitude of the radial oscillations $\lambda = 1-\left(\frac{2T}{\left|U\right|} \right)_{0} $ 
is expressed in terms of the virial ratio at t=0 with ($0\le\lambda\le1$). For $\lambda=0 $ the model reduces to the equilibrium disk of Ref.[8]. In Eq. (1)  ${v}_{r}$ and ${v}_{\bot }$ – denote the radial and tangential components of the velocity of a “particle” with position vector $\vec{r}(x,y)$ whose magnitude is related to the corresponding equilibrium coordinate $r_0$ by $r=\Pi \left(t\right)\, \cdot r_{0} $ . Here $\Omega$ - is the dimensionless parameter [8] characterizing the amplitude of rigid-body rotation of the disk and $\chi$ - is the Heaviside function. The surface density of the nonstationary disk (1) is given by

\beq
    \sigma \left(\vec{r};\, t\right)=\frac{\sigma _{0} }{\Pi^{2} } \sqrt{1-\frac{r^{2} }{R^{2} } }
\eeq
where $\sigma _{0} =\sigma \left(0;\; 0\right)$. In addition in Eq. (1)

\beq 
    {v}_{{a}} =-{\lambda}\frac{{r}\; {\rm sin\psi }}{\sqrt{{\rm 1}-{\rm \lambda}^{{2}} } {\Pi}^{{2}} } ,  \quad \quad  {v}_{{b}} =\frac{{\rm \Omega r}}{{\Pi}^{{2}} }  ,                      
\eeq
where we adopt the standard normalization $\pi ^{2} G\sigma _{0} =2R_{0} $, and the equilibrium disk radius is set to unity throughout.
It should be noted that model (1) undergoes strictly radial oscillations with the period 
\beq
    P\left(\lambda \right)=\frac{2\pi }{\Omega _ 0 \left(1-\lambda ^{2} \right)^{3/2} },       
\eeq
where $\Omega_0=\left(\frac{\pi^2 G\sigma_0}{2R_0}\right)^{1/2}$ is the angular frequency of a particle moving along the nominal (equilibrium) orbit. As noted above we adopt $\Omega_0=1$.

The nonlinear pulsating model (1) like its equilibrium counterpart [8] has an isotropic velocity distribution which represents a somewhat idealized description. Therefore using the well-known averaging procedure with respect to the parameter $\Omega$:
\beq
\Psi_a=
\frac{\displaystyle\int_{-1}^{+1}\rho(\Omega)\,\Psi\,d\Omega}
{\displaystyle\int_{-1}^{+1}\rho(\Omega)\,d\Omega},
\eeq
an anisotropic model can be constructed by considering the weighting function in Eq. (4) for example in the form
\begin{equation}
\rho(\Omega')=
\frac{2}{\pi}\sqrt{1-\Omega'^2}\left(1+\Omega\Omega'\right).
\end{equation}
Then using Eqs. (1)–(3) and (7) in Eq. (6) we obtain the following anisotropic pulsating disk model [26].
\begin{equation}
\Psi_a=
\frac{\sigma_0}{\pi}
\left[
1+\Omega\left(xv_y-yv_x\right)
\right]
\chi(B),
\end{equation}
where the rotation parameter is again represented by $\Omega$, and
$ B=\left(1-\frac{r^2}{\Pi^2}\right)\left(1-\Pi^2v_\perp^2\right)-\Pi^2\left(v_r-v_a\right)^2.
$
In Ref. [26] NADE was also derived in general form for horizontal perturbations against the background of model (8):
\begin{equation}
\begin{aligned}
&a(\psi)=\frac{4}{N(N^2-1)\Pi^3}\gamma_{Nm}\left[(N-1)(N^2+N-\right.\\
&-\left.m^2)\overline{D}_N+i\Omega m(N^2+N-m^2-1)\,d_N\right].
\end{aligned}
\end{equation}
\text{Here}
\begin{equation*}
\begin{aligned}
&\gamma_N^m=\frac{(N+m-1)!!(N-m-1)!!}{(N+m)!!(N-m)!!},\quad \\
&\overline{D}_N=\int_{-\infty}^{\psi}EW^{N-1}\frac{dP_N}{d(\cosh)}\,d\psi_1,\\ &d_N=\int_{-\infty}^{\psi}EI_N d\psi_1,\quad E=\Pi^3(\psi_1)\,S(\psi,\psi_1)\,a(\psi_1),\\
&I_N=W^{N-1}\left[\frac{1}{N+2}\frac{d^2P_{N+1}}{d(\cosh)^2}-\frac{P_N}{d(\cosh)}\right]\tg{h},\\
\text{and}\\
&W=\frac{1+\lambda\cos\psi_1}{1+\lambda\cos\psi},\quad\cosh=\\
&=\frac{(\cos\psi+\lambda)(\cos\psi_1+\lambda)+(1-\lambda^2)\sin\psi\sin\psi_1
}{(1+\lambda\cos\psi)(1+\lambda\cos\psi_1)}.
\end{aligned}
\end{equation*}
Here $P_N(\cosh)$ denotes the Legendre polynomial while $N;m$ --- are the radial and azimuthal wavenumbers respectively. Thus using the NADE (9) we can investigate the behavior of arbitrary types of horizontal perturbation modes against the background of the nonlinear nonstationary model (8).
\section{Instability Analysis of Individual Small-Scale Perturbation Modes}
Based on the NADE (9) we investigated a number of specific small-scale
perturbation modes within the framework of the pulsating anisotropic disk
model: $(m;N)=(12;20)$, $(14;20)$, $(16;20)$ and $(18;20)$.
For example for the case $m=12,\;N=20$, the corresponding NADE is given by:
\beq
\begin{aligned}    
&a_{12;20}(\psi)=\\
&=\frac{79089525}{8589934592\cdot\Pi^3}\left[437\,\overline{D}_{20}+275\,i\,\Omega\,d_{20}
\right].
\end{aligned}
\eeq
After some simplifications we introduce the notation
\[
\begin{aligned}
&\ell_{\tau}(\psi)=\int_{-\infty}^{\psi}\left(1+\lambda\cos\psi_{1}\right)^3S(\psi,\psi_{1})\times\\ &\times a_{12;20}(\psi_{1})\left(\lambda+\cos\psi_{1}\right)^{19-\tau} \sin^{\tau}\psi_{1}\,d\psi_{1}, (\tau=0\text{--}19).
\end{aligned}
\]
The NADE (10) can then be written in the form
\begin{equation}
a_{12;20}(\psi)=\frac{1}{(1+\lambda\cos\psi)^{41}}K_{12;20}(\psi).
\end{equation}
Owing to the cumbersome expression for the function $K_{12;20}(\psi)$ it is presented in the Appendix (Item.1). It should be noted that Eq. (11) is an integral equation which makes it inconvenient for further analysis. Therefore following the approach described for example in Ref. [10], we can transform the NADE into a differential form.
\begin{equation}
\begin{aligned}
&(1+\lambda\cos\psi)\frac{d^2\ell_{\tau}(\psi)}{d\psi^2}+\lambda\sin\psi
\frac{d\ell_{\tau}(\psi)}{d\psi}+\ell_{\tau}(\psi)=\\
&=\frac{(\lambda+\cos\psi)^{19-\tau}\sin^{\tau}\psi}
{(1+\lambda\cos\psi)^{38}}K_{12;20}(\psi).
\end{aligned}
\end{equation}
The NADE (12) constitutes a system of twenty second-order differential equations with two free parameters. Since the unknown functions are complex-valued separating each of them into its real and imaginary parts yields a system of forty equations. Therefore we numerically investigated system (12) using the method of stability of periodic solutions [27].
\begin{figure*}[t]
\centering
\includegraphics[scale=0.9]{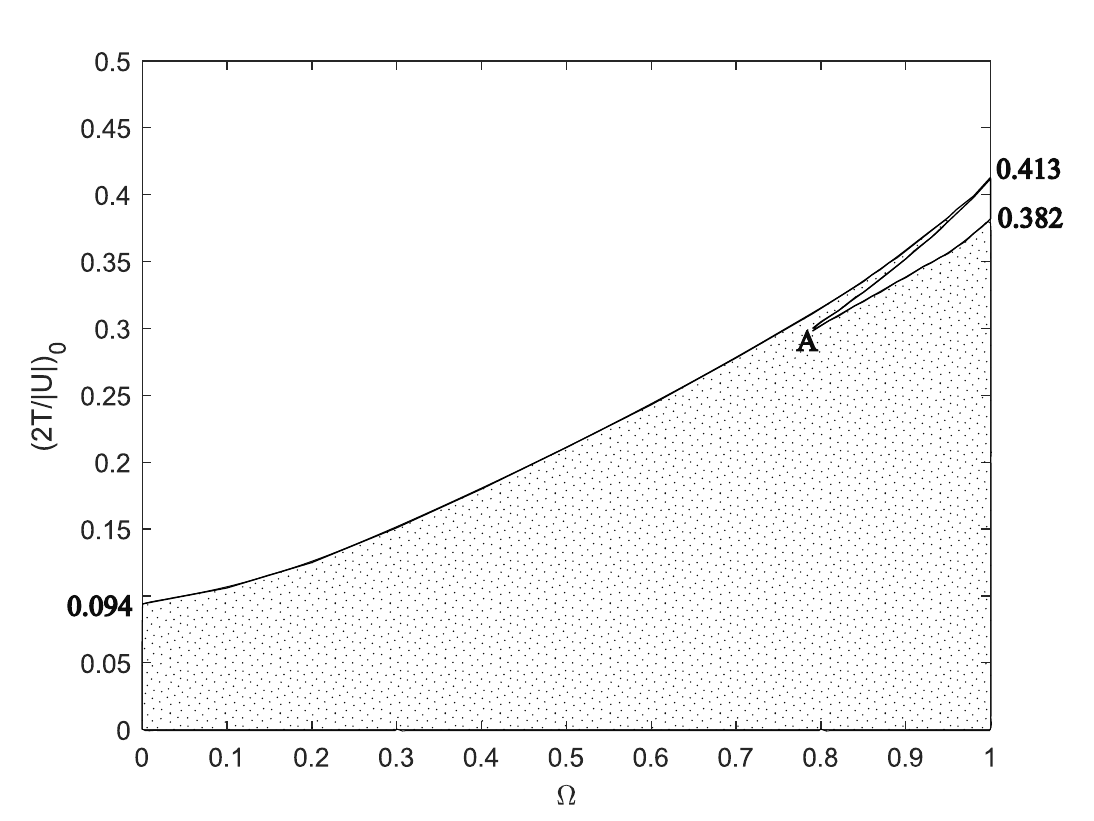}
\caption{\small
    Critical dependence of the initial virial ratio on the rotation parameter
    of disk (8) for the mode $m=12$, $N=20$. Here $A(0.79;0.299)$.
    The unstable region is shaded.}
		\label{fig1}
\end{figure*}
\begin{figure*}[!b]
\centering
\includegraphics[scale=0.90]{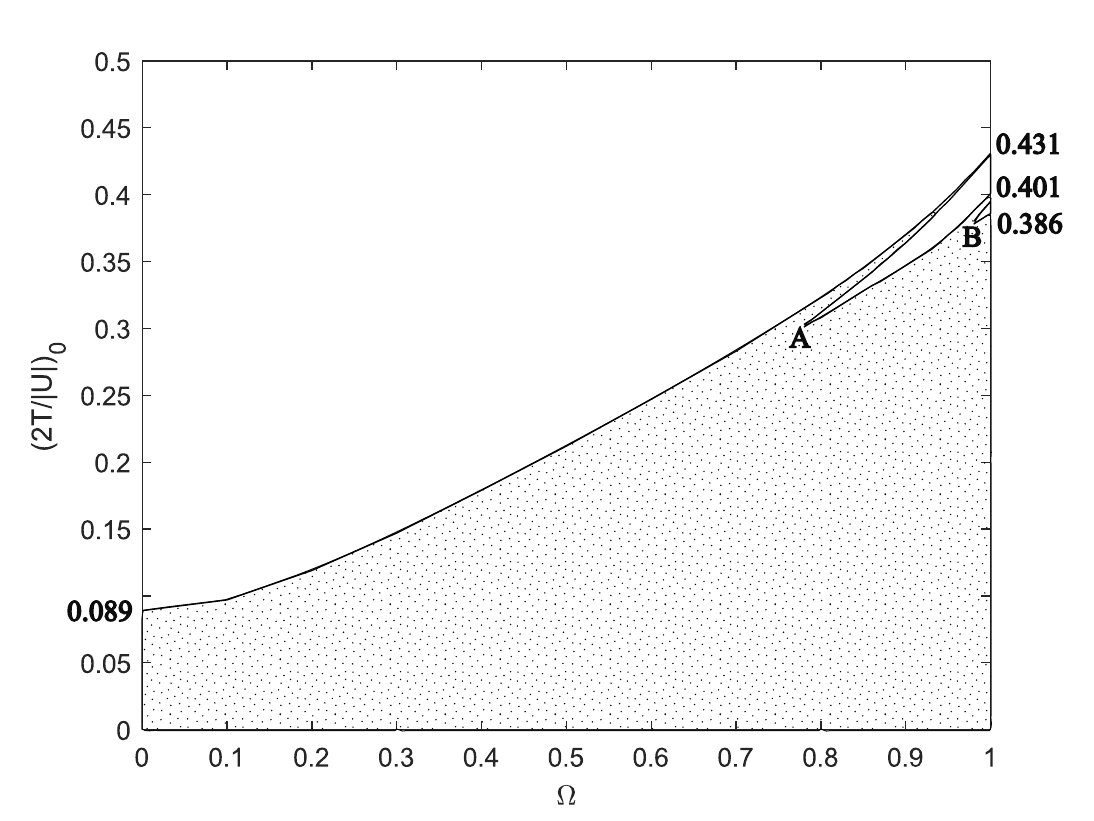}
\caption{\small
    Critical dependence of the initial virial ratio on the rotation parameter
    of model (8) for the mode $m=14$, $N=20$. Here $A(0.78;0.302)$ and $B(0.98;0.379)$. The unstable region is shaded.}
		\label{fig2}
\end{figure*}
\begin{figure*}[!b]
\centering
\includegraphics[scale=0.90]{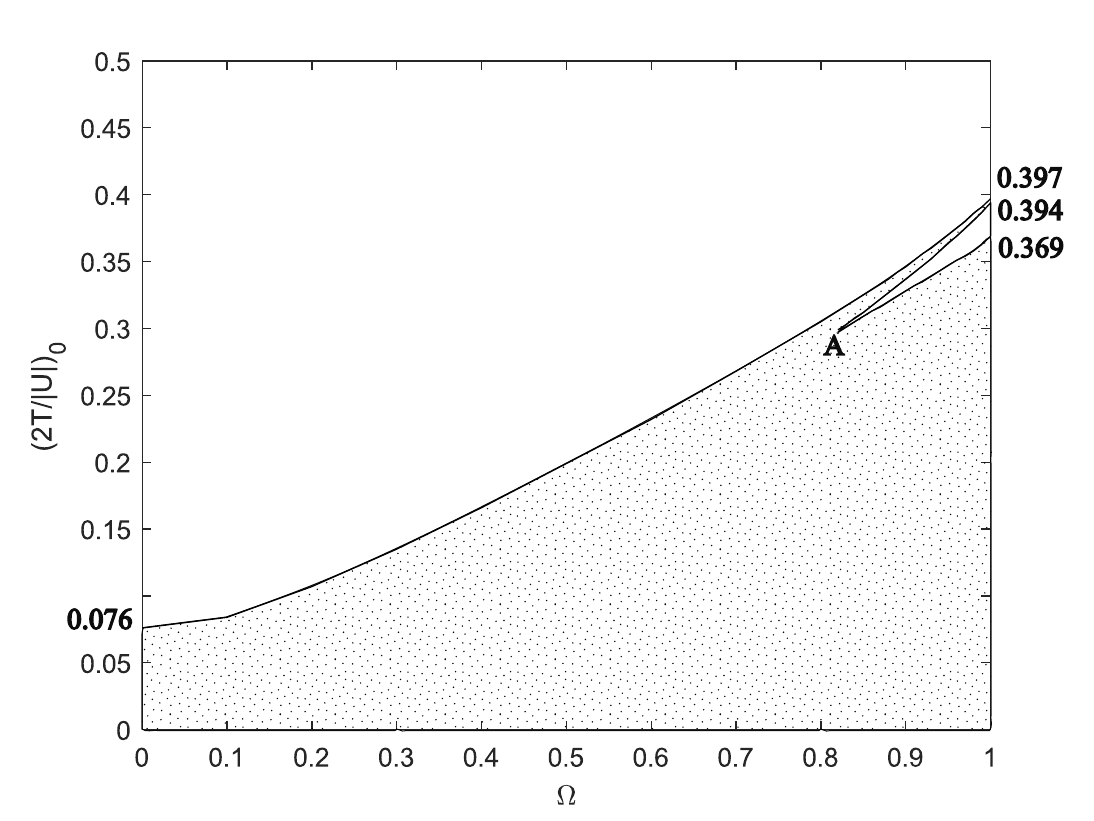}
\caption{\small
    Critical dependence of the initial virial ratio on the rotation parameter
    of disk (8) for the mode $m=16$, $N=20$. Here $A(0.82;0.298)$.
    The unstable region is shaded.}
		\label{fig3}
\end{figure*}
During the calculations by varying the rotation parameter $\Omega$ and the pulsation amplitude $\lambda$ over the range from $0$ to $1$, we determined the critical values of the initial virial ratio $(2T/|U|)_0^{*}$ above which model (8) becomes unstable with respect to the small-scale perturbation mode $(12;20)$. For this purpose the characteristic equation was solved using the solutions of the NADE (12) at $\psi=2\pi$ for different values of the rotation parameter $\Omega$ and the pulsation amplitude $\lambda$ (see [27]). Using the numerical solutions of the NADE (12), we constructed a critical diagram of the initial virial ratio as a function of the rotation parameter of the pulsating anisotropic model (8) (Fig. 1). Analysis of the results shows that the nature of the instability of the mode $(12;20)$ against the background of the non-rotating model (8) differs physically for different values of the initial virial ratio. In the intervals $ 0\le(2T/|U|)_0\le0.024,\quad 0.038\le(2T/|U|)_0\le0.055,\quad 0.075\le(2T/|U|)_0\le0.087$ radial-motion instability occurs which is aperiodic in nature whereas in the intervals
$0.025\le(2T/|U|)_0\le0.037,\quad 0.056\le(2T/|U|)_0\le0.074,\quad 0.088\le(2T/|U|)_0\le0.094 $ an oscillatory-resonant instability occurs. When the anisotropic model (8) is rotating only oscillatory instability is observed.
As can be seen from Fig. 1 with increasing rotation parameter of the anisotropic self-gravitating disk model (8) the main instability region gradually expands starting from an initial virial ratio of approximately $(2T/|U|)_0\approx0.094$
and extending up to $0.413$. At high rotation rates a stable island is observed whose onset is located at point $A(0.79;0.299)$. 

We now investigate the development of the instability of the small-scale mode $\mathbf{m=14;\;N=20}$ against the background of the anisotropic model (8). Based on Eq. (9) we obtain the following NADE in differential form for this mode.
\begin{equation}
\begin{aligned}
&(1+\lambda\cos\psi)\frac{d^2\eta_{\tau}(\psi)}{d\psi^2}+\lambda\sin\psi
\frac{d\eta_{\tau}(\psi)}{d\psi}+\eta_{\tau}(\psi)\\
&=\frac{(\lambda+\cos\psi)^{19-\tau}\sin^{\tau}\psi}
{(1+\lambda\cos\psi)^{38}}K_{14;20}(\psi).
\end{aligned}
\end{equation}
\text{where}
\[
\begin{aligned}
&\eta_{\tau}(\psi)=\int_{-\infty}^{\psi}\left(1+\lambda\cos\psi_1\right)^3
S(\psi,\psi_1)a_{14;20}(\psi_1)\times\\
&\times\left(\lambda+\cos\psi_1\right)^{19-\tau}\sin^{\tau}\psi_1\,d\psi_1,\quad (\tau=0\text{--}19).
\end{aligned}
\]
In this case the expression for $K_{14;20}(\psi)$ is also cumbersome and is therefore given in the Appendix (Item.2).

The results of the numerical calculations of the NADE (13) are presented in the form of a critical diagram of the initial virial ratio as a function of the rotation parameter of model (8) (Fig. 2). For the non-rotating model (8) six alternating instability regions with periodic and aperiodic behavior are also observed. In the intervals
$0\le(2T/|U|)_0\le0.023, 0.044\le(2T/|U|)_0\le0.062,\quad 0.076\le(2T/|U|)_0\le0.084$ we observe oscillatory instability whereas in the ranges $ 0.024\le(2T/|U|)_0\le0.043,\quad 0.063\le(2T/|U|)_0\le0.076,\\ 0.085\le(2T/|U|)_0\le0.089$
radial-orbit instability occurs. When the model is rotating only oscillatory instability is observed and rotation acts predominantly as a destabilizing factor. As the rotation parameter approaches its maximum value two sharply pointed stable islands emerge located at the points $A(0.78;0.302)$ and $B(0.98;$ $0.379)$.

The behavior of the small-scale mode $\mathbf{m=16;}$\\$\mathbf{N=20}$ within the framework of the nonstationary model (8) in accordance with Eq. (9) is described by the following NADE:
\begin{equation}
\begin{aligned}
&(1+\lambda\cos\psi)
\frac{d^2\mu_{\tau}(\psi)}{d\psi^2}+\lambda\sin\psi\frac{d\mu_{\tau}(\psi)}{d\psi}+\mu_{\tau}(\psi)=\\
&=\frac{(\lambda+\cos\psi)^{19-\tau}\sin^{\tau}\psi}
{(1+\lambda\cos\psi)^{38}}K_{16;20}(\psi).
\end{aligned}
\end{equation}

\text{Here}
\[
\begin{aligned}
&\mu_{\tau}(\psi)=\int_{-\infty}^{\psi}\left(1+\lambda\cos\psi_1\right)^3
S(\psi,\psi_1)a_{16;20}(\psi_1)\times\\
&\times\left(\lambda+\cos\psi_1\right)^{19-\tau}
\sin^{\tau}\psi_1\,d\psi_1,\quad (\tau=0\text{--}19).
\end{aligned}
\]

The function $K_{16;20}(\psi)$ is given in the Appendix (Item 3). Analysis of the numerical solutions of the NADE (14) shows that for the non-rotating model (8), the small-scale mode $(16;20)$ exhibits aperiodic instability in the intervals $0\leq(2T/|U|)_0\leq0.027,\quad 0.048\leq(2T/|U|)_0\leq0.061,\quad 0.071\leq(2T/|U|)_0\leq0.076, $
whereas oscillatory instability occurs in the intervals $ 0.028\leq(2T/|U|)_0\leq0.047,\quad 0.062\leq(2T/|U|)_0\leq0.070.$
However when model (8) is rotating only instability of an oscillatory
nature is observed. The critical diagram of the initial virial ratio as
a function of the rotation parameter of the nonstationary model (8) for
this mode shows that as the rotation parameter of the disk increases
the instability region expands, that is the onset of instability occurs
at increasingly larger values of the initial virial ratio (Fig. 3).
In addition at large values of the disk rotation parameter a stable
island develops the onset of which corresponds to the point
$A(0.82;\,0.298)$.

Finally we consider the small-scale mode $\mathbf{m=18;}$\\$\mathbf{N=20}$. Substituting these values into Eq. (9), we obtain the following NADE for this perturbation mode:
\begin{equation}
\begin{aligned}
&(1+\lambda\cos\psi)\frac{d^2\zeta_{\tau}(\psi)}{d\psi^2}+\lambda\sin\psi
\frac{d\zeta_{\tau}(\psi)}{d\psi}+\zeta_{\tau}(\psi)=\\
&=\frac{(\lambda+\cos\psi)^{19-\tau}\sin^{\tau}\psi}
{(1+\lambda\cos\psi)^{38}}K_{18;20}(\psi).
\end{aligned}
\end{equation}
Here
\[
\begin{aligned}
&\zeta_{\tau}(\psi)=\int_{-\infty}^{\psi}\left(1+\lambda\cos\psi_1\right)^3
S(\psi,\psi_1)a_{18;20}(\psi_1)\\
&\left(\lambda+\cos\psi_1\right)^{19-\tau}\sin^{\tau}\psi_1\,d\psi_1,
\quad(\tau=0\text{--}19).
\end{aligned}
\]

The function $K_{18;20}(\psi)$ is given in the Appendix (Item 4).

Here based on the results of the analysis of the NADE (15), we can also
conclude that in the absence of rotation of model (8) oscillatory
instability occurs in the intervals $0\leq(2T/|U|)_0\leq0.026,\quad0.041\leq(2T/|U|)_0\leq0.050,$
whereas aperiodic instability is observed in the ranges
$0.027\leq(2T/|U|)_0\leq0.040,\quad 0.051\leq(2T/|U|)_0\leq0.056.$
However when the model is rotating only oscillatory instability is
observed. The rotation parameter plays a destabilizing role. It should
be noted that in this case no stable island is observed in contrast
to the first three cases (Fig. 4).
\begin{figure*}[t]
\centering
\includegraphics[scale=0.90]{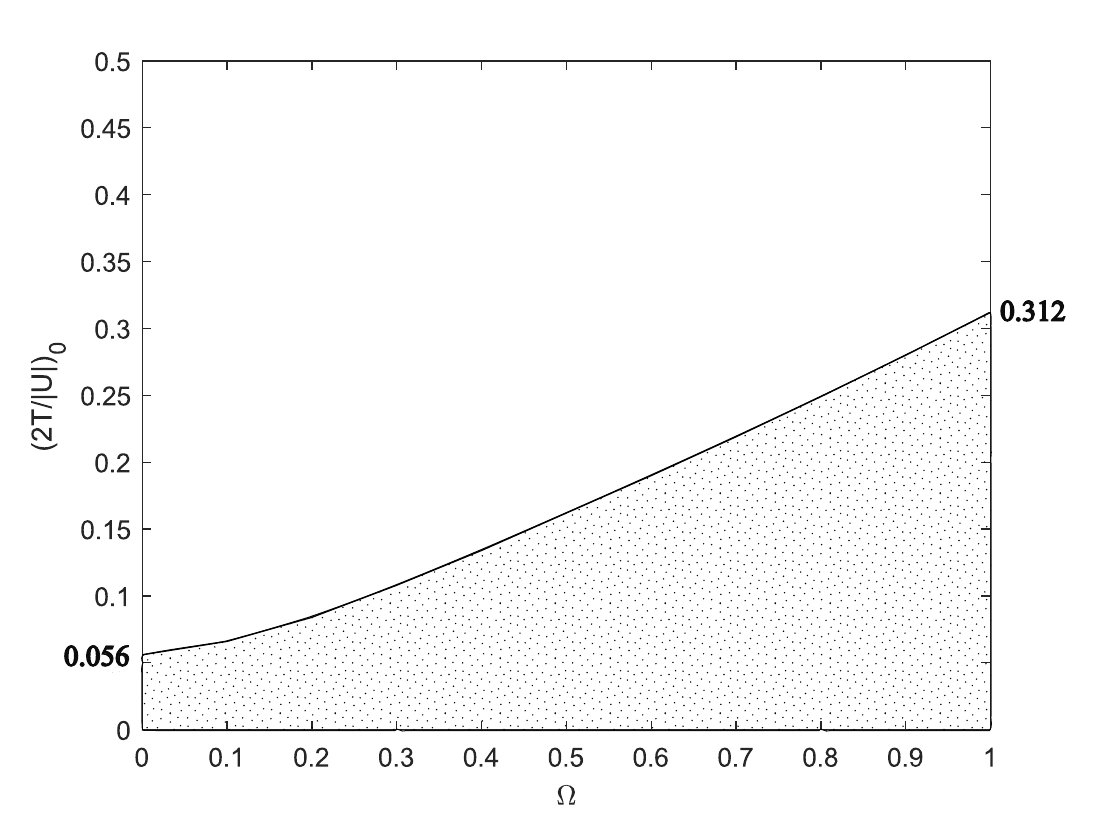}
\caption{\small
    Critical dependence of the initial virial ratio on the rotation parameter of disk (8) for the mode \(m=18,\;N=20\). The unstable region is shaded.}
		\label{fig4}
\end{figure*}
Thus the shaded regions in Figs. 1–4 define the instability regions. Therefore for a fixed \(N\) as the value of \(m\) increases the instability region becomes narrower and the stable island may gradually disappear. As the rotation rate of model (8) increases the instability region expands. 

During the numerical calculations of the NADE for the small-scale horizontal perturbation modes considered here the corresponding instability increments were also calculated for each value of \(\lambda\) using the formula
$$
\mathrm{Incr}=\frac{\ln\left|k_{\max}\right|}{P(\lambda)},
$$
where \(\ln\left|k_{\max}\right|\) is the natural logarithm of the largest value of the modulus of the root of the characteristic equation.
\begin{figure*}[t]
\centering
\includegraphics[width=3.4 in, height=2.5in]{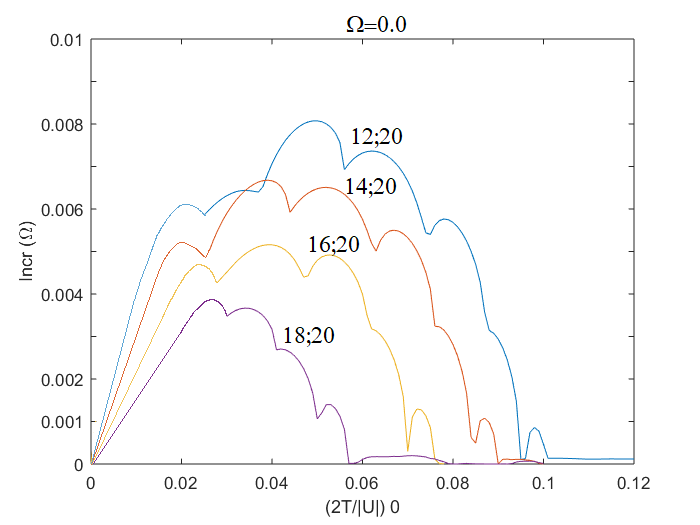} \quad \includegraphics[width=3.4 in, height=2.5in]{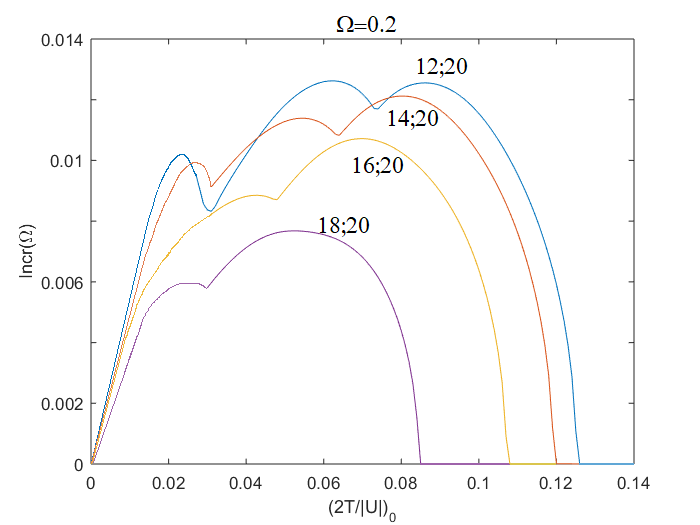} \\
\centering
\includegraphics[width=3.4 in, height=2.5in]{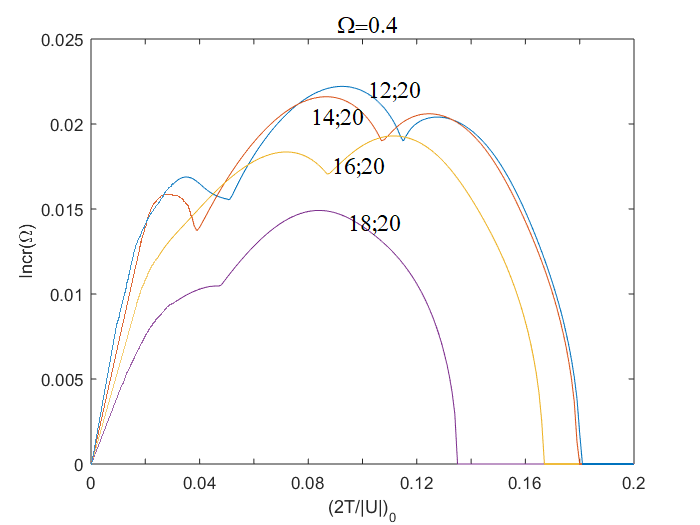} \quad 
\includegraphics[width=3.4 in, height=2.5in]{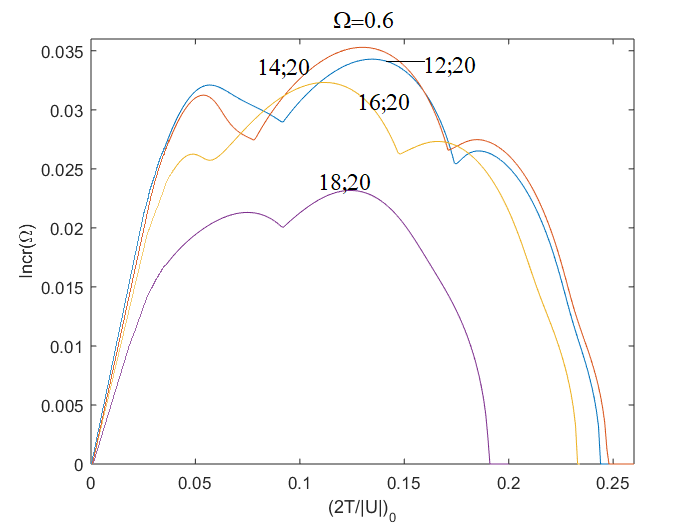} \\ 
\centering
\includegraphics[width=3.4 in, height=2.5in]{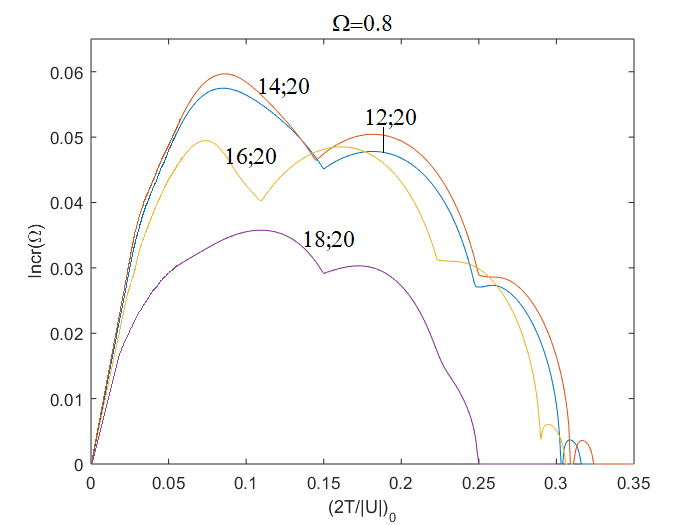} \quad \includegraphics[width=3.4 in, height=2.5in]{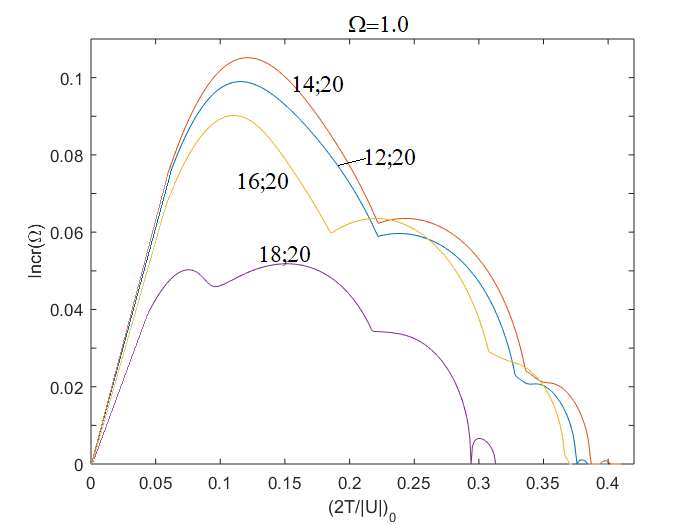} \\
\caption{\small 
		 Comparison of the instability increments of small-scale horizontal perturbation modes for different values of the rotation parameter of model (8). The increment is given in units of \(\Omega_0=\left(\pi^2G\sigma_0/2R_0\right)^{1/2}\).}
		\label{fig5}
\end{figure*}
Figure 5 compares the instability increments of these small-scale modes
for different values of the rotation parameter of disk (8). When the
model is non-rotating the instability-increment curves of the
perturbation modes considered are arranged in descending order of $m$;
that is the instability increment decreases with increasing azimuthal
wavenumber. This behavior of the instability increment is observed only
for rotation parameter values of model (8) in the range
$0\leq\Omega\leq 0.6$. As the rotation parameter approaches its maximum
value the $(14;20)$ mode becomes dominant. It should also be noted
that with increasing rotation parameter of the disk the range of
critical initial virial-ratio values increases.

\section{Conclusion}

\par We summarize the main results obtained in this work as follows:
\begin{enumerate}

\item The NADE for four small-scale horizontal perturbation modes
were derived against the background of the nonlinear pulsating anisotropic
disk model (8) and the results of their numerical calculations were obtained.

\item Critical dependences of the initial virial ratio on the rotation
parameter of the model were constructed for the considered small-scale
perturbation modes and the instability increments were calculated for
different values of the model parameters.

\item The characteristic timescales for the development of small-scale
disk structures and the critical values of the initial virial ratio were
determined as functions of the main physical parameters of disk model (8).

\item It was established that for the non-rotating model (8) both
oscillatory and aperiodic instabilities of small-scale horizontal
perturbation modes occur at different values of the initial virial ratio.
When the model is rotating however only oscillatory instability is observed.

\item It was shown that as the rotation rate of the pulsating
self-gravitating disk model (8) increases the instability region of the
small-scale horizontal perturbation modes systematically expands.
Furthermore for a fixed value of the principal perturbation index $N$
the instability region becomes narrower with increasing azimuthal
wavenumber $m$ and the stable island may gradually disappear.

\item It was found that for rotation parameter values
$0\leq\Omega\leq 0.6$, the instability increment of the modes decreases with
increasing azimuthal wavenumber $m$. As the rotation parameter approaches
its maximum value the $(14;20)$ mode becomes dominant. It was also found
that increasing the rotation parameter of the disk leads to an expansion
of the range of critical initial virial-ratio values.
\end{enumerate}

The authors are grateful to the reviewer for the valuable comments.
We also express our gratitude to the Academy of Sciences of the Republic
of Uzbekistan and the Agency for Innovation of the Republic of Uzbekistan
for their financial support.

\Funding {This work was supported by the state budget of the Republic of Uzbekistan.}

\onecolumn
\appendix
\section*{Appendix}
\[\textbf{Appendix (Item 1). Expression for the Function}K_{12;20}\textbf{:}\]
\allowdisplaybreaks
\begin{gather*}
K_{12;20}(\psi)={}\frac{30705345}{120259084288}(779((\frac{43648605}{128}c^{7}e^{12}b^{12}+210c^{19}+\frac{72747675}{16384}c^{3}e^{16}b^{16}-\frac{101846745}{128}c^{9}e^{10}b^{10}+\\
+101745c^{15}e^{4}b^{4}+\frac{27776385}{32}c^{11}e^{8}b^{8}-\frac{17955}{2}c^{17}e^{2}b^{2}-\frac{3561075}{8}c^{13}e^{6}b^{6}-\frac{130945815}{2048}c^{5}e^{14}b^{14}-\frac{4849845}{65536}ce^{18}b^{18})l_{0}(\psi)
+\\
+(-\frac{101846745}{8}e^{8}b^{7}c^{12}+\frac{16787925}{4}e^{6}b^{5}c^{14}+\frac{1120314195}{64}c^{10}e^{10}b^{9}+\frac{480134655}{32768}e^{18}b^{17}c^{2}
-\frac{800224425}{2048}e^{16}b^{15}c^{4}+\\
+21945e^{2}bc^{18}+\frac{3360942585}{1024}e^{14}b^{13}c^{6}-\frac{4849845}{65536}e^{20}b^{19}-\frac{1440403965}{128}c^{8}e^{12}b^{11}-\frac{1119195}{2}e^{4}b^{3}c^{16})l_{1}(\psi)+ \\
+(951615c^{17}e^{4}b^{2}+\frac{24923353455}{2048}e^{16}b^{14}c^{5}-\frac{129243519405}{2048}e^{14}b^{12}c^{7}+\frac{567431865}{4}c^{9}e^{12}b^{10}-\frac{17955}{2}c^{19}e^{2}-\\
-\frac{18804612645}{128}c^{11}e^{10}b^{8}-\frac{112021245}{8}e^{6}b^{4}c^{15}+\frac{557664345}{8}c^{13}e^{8}b^{6}-\frac{56612240685}{65536}e^{18}b^{16}c^{3}+\frac{480134655}{32768}e^{20}b^{18}c)l_{2}(\psi)+\\
+(\frac{41410215}{2}c^{16}e^{6}b^{3}+\frac{92345898645}{4096}c^{4}e^{18}b^{15}-\frac{377545883715}{2048}e^{16}b^{13}c^{6}+\frac{312567660405}{512}e^{14}b^{11}c^{8}-\\
-\frac{116032541625}{128}c^{10}e^{12}b^{9}+\frac{9879134265}{16}c^{12}e^{10}b^{7}-\frac{1473979815}{8}e^{8}b^{5}c^{14}
-\frac{1119195}{2}e^{4}bc^{18}-\frac{56612240685}{65536}e^{20}b^{17}c^{2}+\\
+\frac{72747675}{16384}e^{22}b^{19})l_{3}(\psi)+(\frac{4057895835}{16}c^{15}e^{8}b^{4}-\frac{5068345066785}{16384}c^{5}e^{18}b^{14}
+\frac{6356560895685}{4096}e^{16}b^{12}c^{7}-\\
-\frac{6844814191215}{2048}e^{14}b^{10}c^{9}+\frac{92345898645}{4096}c^{3}e^{20}b^{16}+\frac{418252837275}{128}c^{11}e^{12}b^{8}+101745c^{19}e^{4}
-\frac{91464786945}{64}c^{13}e^{10}b^{6}-\\
-\frac{112021245}{8}e^{6}b^{2}c^{17}-\frac{800224425}{2048}e^{22}b^{18}c)l_{4}(\psi)+(\frac{60359305545}{32}e^{10}b^{5}c^{14}-\frac{5068345066785}{16384}c^{4}e^{20}b^{15}+ \\
+\frac{20061466289865}{8192}c^{6}e^{18}b^{13}-\frac{15947672565825}{2048}e^{16}b^{11}c^{8}+\frac{24923353455}{2048}e^{22}b^{17}c^{2}+\frac{11235631061655}{1024}e^{14}b^{9}c^{10}- \\
-\frac{445957797285}{64}c^{12}e^{12}b^{7}-\frac{1473979815}{8}e^{8}b^{3}c^{16}+\frac{16787925}{4}e^{6}bc^{18}-\frac{130945815}{2048}e^{24}b^{19})l_{5}(\psi)+
(\frac{71662361085}{8}c^{13}e^{12}b^{6}-\\
-\frac{377545883715}{2048}e^{22}b^{16}c^{3}+\frac{20061466289865}{8192}c^{5}e^{20}b^{14}-\frac{193475919001365}{16384}c^{7}e^{18}b^{12}+\frac{3360942585}{1024}e^{24}b^{18}c+ \\
+\frac{49577467764825}{2048}e^{16}b^{10}c^{9}-\frac{45518078720205}{2048}e^{14}b^{8}c^{11}-\frac{91464786945}{64}c^{15}e^{10}b^{4}+\frac{557664345}{8}c^{17}e^{8}b^{2}-\\
-\frac{3561075}{8}c^{19}e^{6})l_{6}(\psi)+(\frac{43648605}{128}e^{26}b^{19}+\frac{7193712040515}{256}c^{12}e^{14}b^{7}-\frac{129243519405}{2048}e^{24}b^{17}c^{2}+\frac{6356560895685}{4096}e^{22}b^{15}c^{4}- \\
-\frac{193475919001365}{16384}c^{6}e^{20}b^{13}+\frac{146612477846835}{4096}c^{8}e^{18}b^{11}-\frac{97410839120595}{2048}e^{16}b^{9}c^{10}-\frac{445957797285}{64}c^{14}e^{12}b^{5}+ \\
+\frac{9879134265}{16}c^{16}e^{10}b^{3}-\frac{101846745}{8}e^{8}bc^{18})l_{7}(\psi)+(\frac{487779442306065}{8192}c^{11}e^{16}b^{8}+\frac{312567660405}{512}e^{24}b^{16}c^{3}- \\
-\frac{15947672565825}{2048}e^{22}b^{14}c^{5}+\frac{146612477846835}{4096}c^{7}e^{20}b^{12}-\frac{1440403965}{128}e^{26}b^{18}c-\frac{2271939876841635}{32768}c^{9}e^{18}b^{10}- \\
-\frac{45518078720205}{2048}e^{14}b^{6}c^{13}+\frac{418252837275}{128}c^{15}e^{12}b^{4}-\frac{18804612645}{128}c^{17}e^{10}b^{2}+\frac{27776385}{32}c^{19}e^{8})l_{8}(\psi)+ \\
+(\frac{1415489617707165}{16384}c^{10}e^{18}b^{9}+\frac{567431865}{4}e^{26}b^{17}c^{2}-\frac{6844814191215}{2048}e^{24}b^{15}c^{4}+\frac{49577467764825}{2048}e^{22}b^{13}c^{6}- \\
-\frac{2271939876841635}{32768}c^{8}e^{20}b^{11}-\frac{97410839120595}{2048}e^{16}b^{7}c^{12}+\frac{11235631061655}{1024}e^{14}b^{5}c^{14}-\frac{116032541625}{128}c^{16}e^{12}b^{3}- \\
-\frac{101846745}{128}e^{28}b^{19}+\frac{1120314195}{64}c^{18}e^{10}b)l_{9}(\psi)+(\frac{1415489617707165}{16384}c^{9}e^{20}b^{10}-\frac{116032541625}{128}e^{26}b^{16}c^{3}+ \\
+\frac{11235631061655}{1024}e^{24}b^{14}c^{5}-\frac{97410839120595}{2048}e^{22}b^{12}c^{7}-\frac{2271939876841635}{32768}c^{11}e^{18}b^{8}+\frac{49577467764825}{2048}e^{16}b^{6}c^{13}- \\
-\frac{6844814191215}{2048}e^{14}b^{4}c^{15}+\frac{567431865}{4}c^{17}e^{12}b^{2}-\frac{101846745}{128}c^{19}e^{10}+\frac{1120314195}{64}e^{28}b^{18}c)l_{10}(\psi)
\\
+(\frac{487779442306065}{8192}c^{8}e^{22}b^{10}+\frac{418252837275}{128}e^{26}b^{15}c^{4}-\frac{45518078720205}{2048}e^{24}b^{13}c^{6}- \\
-\frac{2271939876841635}{32768}c^{10}e^{20}b^{9}-\frac{1440403965}{128}c^{18}e^{12}b+\frac{146612477846835}{4096}c^{12}e^{18}b^{7}-\frac{15947672565825}{2048}e^{16}b^{5}c^{14}+ \\
+\frac{312567660405}{512}e^{14}b^{3}c^{16}-\frac{18804612645}{128}e^{28}b^{17}c^{2}+\frac{27776385}{32}e^{30}b^{19})l_{11}(\psi)+(\frac{43648605}{128}c^{19}e^{12}+ \\
+\frac{7193712040515}{256}c^{7}e^{24}b^{12}-\frac{445957797285}{64}e^{26}b^{14}c^{5}-\frac{97410839120595}{2048}e^{22}b^{10}c^{9}+\frac{146612477846835}{4096}c^{11}e^{20}b^{8}- \\
-\frac{193475919001365}{16384}c^{13}e^{18}b^{6}+\frac{6356560895685}{4096}e^{16}b^{4}c^{15}-\frac{129243519405}{2048}e^{14}b^{2}c^{17}+\frac{9879134265}{16}e^{28}b^{16}c^{3}-\\
-\frac{101846745}{8}e^{30}b^{18}c)l_{12}(\psi)+(\frac{71662361085}{8}c^{6}e^{26}b^{13}-\frac{45518078720205}{2048}e^{24}b^{11}c^{8}+\frac{49577467764825}{2048}e^{22}b^{9}c^{10}+\\
+\frac{3360942585}{1024}e^{14}bc^{18}-\frac{193475919001365}{16384}c^{12}e^{20}b^{7}+\frac{20061466289865}{8192}c^{14}e^{18}b^{5}-\frac{377545883715}{2048}e^{16}b^{3}c^{16}-\\
-\frac{91464786945}{64}e^{28}b^{15}c^{4}+\frac{557664345}{8}c^{2}e^{30}b^{17}-\frac{3561075}{8}e^{32}b^{19})l_{13}(\psi)+(\frac{60359305545}{32}e^{28}b^{14}c^{5}-\frac{445957797285}{64}e^{26}b^{12}c^{7}+ \\
+\frac{11235631061655}{1024}e^{24}b^{10}c^{9}+\frac{24923353455}{2048}e^{16}b^{2}c^{17}-\frac{15947672565825}{2048}e^{22}b^{8}c^{11}+\frac{20061466289865}{8192}c^{13}e^{20}b^{6}- \\
-\frac{5068345066785}{16384}c^{15}e^{18}b^{4}-\frac{1473979815}{8}e^{30}b^{16}c^{3}+\frac{16787925}{4}e^{32}b^{18}c-\frac{130945815}{2048}c^{19}e^{14})l_{14}(\psi)+\\
+(\frac{4057895835}{16}c^{4}e^{30}b^{15}+\frac{418252837275}{128}e^{26}b^{11}c^{8}+\frac{92345898645}{4096}c^{16}e^{18}b^{3}-\frac{6844814191215}{2048}e^{24}b^{9}c^{10}+ \\
+\frac{6356560895685}{4096}e^{22}b^{7}c^{12}-\frac{5068345066785}{16384}c^{14}e^{20}b^{5}-\frac{91464786945}{64}e^{28}b^{13}c^{6}-\frac{112021245}{8}e^{32}b^{17}c^{2}+101745e^{34}b^{19}
-\\
-\frac{800224425}{2048}e^{16}bc^{18})l_{15}(\psi)+(\frac{41410215}{2}c^{3}e^{32}b^{16}-\frac{116032541625}{128}e^{26}b^{10}c^{9}
+\frac{312567660405}{512}e^{24}b^{8}c^{11}-\\
-\frac{377545883715}{2048}e^{22}b^{6}c^{13}+\frac{92345898645}{4096}c^{15}e^{20}b^{4}+\frac{9879134265}{16}e^{28}b^{12}c^{7}-\frac{1119195}{2}e^{34}b^{18}c
-\frac{1473979815}{8}e^{30}b^{14}c^{5}-\\
-\frac{56612240685}{65536}e^{18}b^{2}c^{17}+\frac{72747675}{16384}c^{19}e^{16})l_{16}(\psi)
+(951615c^{2}e^{34}b^{17}+\frac{567431865}{4}e^{26}b^{9}c^{10}-\frac{129243519405}{2048}e^{24}b^{7}c^{12}+ \\
+\frac{24923353455}{2048}e^{22}b^{5}c^{14}-\frac{18804612645}{128}e^{28}b^{11}c^{8}-\frac{17955}{2}e^{36}b^{19}-\frac{112021245}{8}e^{32}b^{15}c^{4}
+\frac{557664345}{8}c^{6}e^{30}b^{13}-\\
-\frac{56612240685}{65536}e^{20}b^{3}c^{16}+\frac{480134655}{32768}e^{18}bc^{18})l_{17}(\psi)+(\frac{16787925}{4}e^{32}b^{14}c^{5}-\frac{101846745}{8}e^{30}b^{12}c^{7}+ \\
+\frac{3360942585}{1024}e^{24}b^{6}c^{13}-\frac{800224425}{2048}e^{22}b^{4}c^{15}+\frac{1120314195}{64}e^{28}b^{10}c^{9}
+\frac{480134655}{32768}e^{20}b^{2}c^{17}+21945e^{36}b^{18}c-\\
-\frac{1440403965}{128}e^{26}b^{8}c^{11}-\frac{4849845}{65536}c^{19}e^{18}-\frac{1119195}{2}e^{34}b^{16}c^{3})l_{18}(\psi)+(210e^{38}b^{19}
-\frac{101846745}{128}e^{28}b^{9}c^{10}+\\
+101745e^{34}b^{15}c^{4}+\frac{27776385}{32}e^{30}b^{11}c^{8}
-\frac{17955}{2}e^{36}b^{17}c^{2}-\frac{3561075}{8}e^{32}b^{13}c^{6}-\frac{130945815}{2048}e^{24}b^{5}c^{14}+ \\
+\frac{43648605}{128}e^{26}b^{7}c^{12}-\frac{4849845}{65536}e^{20}bc^{18}+\frac{72747675}{16384}e^{22}b^{3}c^{16})l_{19}(\psi)
+2608i\Omega e((-\frac{305540235}{2048}e^{12}b^{13}c^{6}-\frac{1995}{2}bc^{18}-\\
-\frac{43648605}{65536}e^{16}b^{17}c^{2}
+\frac{130945815}{256}e^{10}b^{11}c^{8}-\frac{1526175}{8}e^{4}b^{5}c^{14}-\frac{101846745}{128}e^{8}b^{9}c^{10}+\frac{101745}{4}e^{2}b^{3}c^{16}+\\
+\frac{9258795}{16}e^{6}b^{7}c^{12}+\frac{72747675}{4096}e^{14}b^{15}c^{4}
+\frac{440895}{131072}e^{18}b^{19})l_{0}(\psi)
+(-\frac{107544465}{16}c^{13}e^{6}b^{6}+\frac{1805465025}{128}c^{11}e^{8}b^{8}- \\
-\frac{3477338865}{256}c^{9}e^{10}b^{10}+\frac{5397877485}{65536}c^{3}e^{16}b^{16}
-\frac{182971425}{131072}ce^{18}b^{18}+\frac{1995}{2}c^{19}-\frac{377055}{4}c^{17}e^{2}b^{2}- \\
-\frac{4757697945}{4096}c^{5}e^{14}b^{14}+\frac{12352555215}{2048}c^{7}e^{12}b^{12}+\frac{10886715}{8}c^{15}e^{4}b^{4})l_{1}(\psi)
+(-\frac{13125105}{4}e^{4}b^{3}c^{16}+\frac{377055}{4}e^{2}bc^{18}-\\
-\frac{94673824245}{1024}c^{8}e^{12}b^{11}
+\frac{113966507655}{4096}e^{14}b^{13}c^{6}-\frac{27804161385}{8192}e^{16}b^{15}c^{4}
+\frac{35311721445}{256}c^{10}e^{10}b^{9}+\\
+\frac{456936795}{16}e^{6}b^{5}c^{14}-\frac{3027625965}{32}e^{8}b^{7}c^{12}
-\frac{43648605}{65536}e^{20}b^{19}+\frac{17011051785}{131072}e^{18}b^{17}c^{2})l_{2}(\psi)+ \\
+(-\frac{910922985}{16}e^{6}b^{4}c^{15}+\frac{13125105}{4}c^{17}e^{4}b^{2}-\frac{1355478329205}{4096}e^{14}b^{12}c^{7}
+\frac{734504175405}{1024}c^{9}e^{12}b^{10}-\frac{624776873175}{131072}e^{18}b^{16}c^{3}+ \\
+\frac{537794462205}{8192}e^{16}b^{14}c^{5}-\frac{181153614915}{256}c^{11}e^{10}b^{8}+
+\frac{10038670425}{32}c^{13}e^{8}b^{6}-\frac{101745}{4}c^{19}e^{2}+\frac{5397877485}{65536}e^{20}b^{18}c)l_{3}(\psi)+ \\
+(-\frac{36011930535}{64}e^{8}b^{5}c^{14}+\frac{910922985}{16}c^{16}e^{6}b^{3}
-\frac{6457301876025}{2048}c^{10}e^{12}b^{9}+\frac{9067517554095}{4096}e^{14}b^{11}c^{8}- \\
-\frac{11315478909735}{16384}e^{16}b^{13}c^{6}+\frac{2841078440655}{32768}c^{4}e^{18}b^{15}
-\frac{27804161385}{8192}e^{20}b^{17}c^{2}+\frac{260186689035}{128}c^{12}e^{10}b^{7}-\\
-\frac{10886715}{8}e^{4}bc^{18}+\frac{72747675}{4096}e^{22}b^{19})l_{4}(\psi)
+(\frac{36011930535}{64}c^{15}e^{8}b^{4}
-\frac{437433194205}{128}c^{13}e^{10}b^{6}+\frac{16976779693965}{2048}c^{11}e^{12}b^{8}- \\
-\frac{36376092677925}{4096}e^{14}b^{10}c^{9}+\frac{70102234897695}{16384}e^{16}b^{12}c^{7}
-\frac{28796615589105}{32768}c^{5}e^{18}b^{14}+\frac{537794462205}{8192}c^{3}e^{20}b^{16}- \\
-\frac{4757697945}{4096}e^{22}b^{18}c-\frac{456936795}{16}e^{6}b^{2}c^{17}+\frac{1526175}{8}c^{19}e^{4})l_{5}(\psi)
+(-\frac{6848147357415}{512}c^{12}e^{12}b^{7}+\frac{437433194205}{128}e^{10}b^{5}c^{14}- \\
-\frac{305540235}{2048}e^{24}b^{19}+\frac{90588503960955}{4096}e^{14}b^{9}c^{10}
-\frac{133892925160995}{8192}e^{16}b^{11}c^{8}+\frac{174186589920525}{32768}c^{6}e^{18}b^{13}- \\
-\frac{11315478909735}{16384}c^{4}e^{20}b^{15}+\frac{113966507655}{4096}e^{22}b^{17}c^{2}
-\frac{10038670425}{32}e^{8}b^{3}c^{16}+\frac{107544465}{16}e^{6}bc^{18})l_{6}(\psi)+ \\
+(-\frac{142481264361345}{4096}e^{14}b^{8}c^{11}+\frac{6848147357415}{512}c^{13}e^{12}b^{6}
+\frac{323723094429735}{8192}e^{16}b^{10}c^{9}-\frac{655162385283315}{32768}c^{7}e^{18}b^{12}+ \\
+\frac{70102234897695}{16384}c^{5}e^{20}b^{14}-\frac{1355478329205}{4096}e^{22}b^{16}c^{3}
+\frac{12352555215}{2048}e^{24}b^{18}c-\frac{260186689035}{128}c^{15}e^{10}b^{4}-\frac{9258795}{16}c^{19}e^{6}+ \\
+\frac{3027625965}{32}c^{17}e^{8}b^{2})l_{7}(\psi)
+(-\frac{2008258828631055}{32768}e^{16}b^{9}c^{10}
+\frac{142481264361345}{4096}c^{12}e^{14}b^{7}-\frac{16976779693965}{2048}c^{14}e^{12}b^{5}+ \\
+\frac{3138386978624625}{65536}c^{8}e^{18}b^{11}-\frac{133892925160995}{8192}c^{6}e^{20}b^{13}
+\frac{9067517554095}{4096}e^{22}b^{15}c^{4}-\frac{94673824245}{1024}e^{24}b^{17}c^{2}+\\
+\frac{181153614915}{256}c^{16}e^{10}b^{3}+\frac{130945815}{256}e^{26}b^{19}-\frac{1805465025}{128}e^{8}bc^{18})l_{8}(\psi)
+(-\frac{4845557363581935}{65536}c^{9}e^{18}b^{10}+\\
+\frac{2008258828631055}{32768}c^{11}e^{16}b^{8}
+\frac{6457301876025}{2048}c^{15}e^{12}b^{4}-\frac{90588503960955}{4096}e^{14}b^{6}c^{13}
+\frac{323723094429735}{8192}c^{7}e^{20}b^{12}-\\
-\frac{36376092677925}{4096}e^{22}b^{14}c^{5}
+\frac{734504175405}{1024}e^{24}b^{16}c^{3}-\frac{35311721445}{256}c^{17}e^{10}b^{2}
-\frac{3477338865}{256}e^{26}b^{18}c+\\
+\frac{101846745}{128}c^{19}e^{8})l_{9}(\psi)
+(-\frac{2008258828631055}{32768}c^{8}e^{20}b^{11}+\frac{4845557363581935}{65536}c^{10}e^{18}b^{9}
-\frac{734504175405}{1024}c^{1}e^{12}b^{3}+\\
+\frac{36376092677925}{4096}e^{14}b^{5}c^{14}
-\frac{323723094429735}{8192}e^{16}b^{7}c^{12}+\frac{90588503960955}{4096}e^{22}b^{13}c^{6}
-\frac{6457301876025}{2048}e^{24}b^{15}c^{4}+\\
+\frac{3477338865}{256}c^{18}e^{10}b
+\frac{35311721445}{256}e^{26}b^{17}c^{2}-\frac{101846745}{128}e^{28}b^{19})l_{10}(\psi)
+(-\frac{142481264361345}{4096}e^{22}b^{12}c^{7}+\\
+\frac{2008258828631055}{32768}c^{9}e^{20}b^{10}
+\frac{94673824245}{1024}c^{17}e^{12}b^{2}-\frac{9067517554095}{4096}e^{14}b^{4}c^{15}
+\frac{133892925160995}{8192}e^{16}b^{6}c^{13}-\\
-\frac{3138386978624625}{65536}c^{11}e^{18}b^{8}
+\frac{16976779693965}{2048}e^{24}b^{14}c^{5}-\frac{181153614915}{256}e^{26}b^{16}c^{3}
-\frac{130945815}{256}c^{19}e^{10}+\\
+\frac{1805465025}{128}e^{28}b^{18}c)l_{11}(\psi)
+(-\frac{6848147357415}{512}e^{24}b^{13}c^{6}+\frac{142481264361345}{4096}c^{8}e^{22}b^{11}
-\frac{12352555215}{2048}c^{18}e^{12}b+\\
+\frac{1355478329205}{4096}e^{14}b^{3}c^{16}
-\frac{70102234897695}{16384}e^{16}b^{5}c^{14}+\frac{655162385283315}{32768}c^{12}e^{18}b^{7}
-\frac{323723094429735}{8192}c^{10}e^{20}b^{9}+\\
+\frac{260186689035}{128}e^{26}b^{15}c^{4}
+\frac{9258795}{16}e^{30}b^{19}-\frac{3027625965}{32}e^{28}b^{17}c^{2})l_{12}(\psi)
+(-\frac{437433194205}{128}e^{26}b^{14}c^{5}+\\
+\frac{6848147357415}{512}c^{7}e^{24}b^{12}
+\frac{305540235}{2048}c^{19}e^{12}-\frac{113966507655}{4096}e^{14}b^{2}c^{17}
+\frac{11315478909735}{16384}e^{16}b^{4}c^{15}-\\
-\frac{174186589920525}{32768}c^{13}e^{18}b^{6}
+\frac{133892925160995}{8192}c^{11}e^{20}b^{8}-\frac{90588503960955}{4096}e^{22}b^{10}c^{9}
+\frac{10038670425}{32}e^{28}b^{16}c^{3}-\\
-\frac{107544465}{16}e^{30}b^{18}c)l_{13}(\psi)
+(\frac{437433194205}{128}c^{6}e^{26}b^{13}-\frac{36011930535}{64}e^{28}b^{15}c^{4}
+\frac{4757697945}{4096}e^{14}bc^{18}-\\
-\frac{537794462205}{8192}e^{16}b^{3}c^{16}
+\frac{28796615589105}{32768}c^{14}e^{18}b^{5}-\frac{70102234897695}{16384}c^{12}e^{20}b^{7}
+\frac{36376092677925}{4096}e^{22}b^{9}c^{10}-\\
-\frac{16976779693965}{2048}e^{24}b^{11}c^{8}
+\frac{456936795}{16}c^{2}e^{30}b^{17}-\frac{1526175}{8}e^{32}b^{19})l_{14}(\psi)
+(-\frac{910922985}{16}e^{30}b^{16}c^{3}+ \\
+\frac{36011930535}{64}e^{28}b^{14}c^{5}+\frac{27804161385}{8192}e^{16}b^{2}c^{17}
-\frac{2841078440655}{32768}c^{15}e^{18}b^{4}+\frac{11315478909735}{16384}c^{13}e^{20}b^{6}-\\
-\frac{9067517554095}{4096}e^{22}b^{8}c^{11}+\frac{6457301876025}{2048}e^{24}b^{10}c^{9}
-\frac{260186689035}{128}e^{26}b^{12}c^{7}+\frac{10886715}{8}e^{32}b^{18}c-\\
-\frac{72747675}{4096}c^{19}e^{14})l_{15}(\psi)
+(-\frac{13125105}{4}e^{32}b^{17}c^{2}+\frac{910922985}{16}c^{4}e^{30}b^{15}-\frac{537794462205}{8192}c^{14}e^{20}b^{5}+\\
+\frac{624776873175}{131072}c^{16}e^{18}b^{3}-\frac{734504175405}{1024}e^{24}b^{9}c^{10}
+\frac{1355478329205}{4096}e^{22}b^{7}c^{12}+\frac{181153614915}{256}e^{26}b^{11}c^{8}- \\
-\frac{10038670425}{32}e^{28}b^{13}c^{6}+\frac{101745}{4}e^{34}b^{19}-\frac{5397877485}{65536}e^{16}bc^{18})l_{16}(\psi)
+(-\frac{377055}{4}e^{34}b^{18}c+\frac{13125105}{4}c^{3}e^{32}b^{16}+\\
+\frac{3027625965}{32}e^{28}b^{12}c^{7}
+\frac{27804161385}{8192}c^{15}e^{20}b^{4}-\frac{113966507655}{4096}e^{22}b^{6}c^{13}
+\frac{94673824245}{1024}e^{24}b^{8}c^{11}-\\
-\frac{35311721445}{256}e^{26}b^{10}c^{9}
-\frac{456936795}{16}e^{30}b^{14}c^{5}+\frac{43648605}{65536}c^{19}e^{16}
-\frac{17011051785}{131072}e^{18}b^{2}c^{17})l_{17}(\psi)+\\
+(-\frac{1805465025}{128}e^{28}b^{11}c^{8}
+\frac{107544465}{16}c^{6}e^{30}b^{13}-\frac{5397877485}{65536}e^{20}b^{3}c^{16}+\frac{4757697945}{4096}e^{22}b^{5}c^{14}+ \\
+\frac{3477338865}{256}e^{26}b^{9}c^{10}+\frac{182971425}{131072}e^{18}bc^{18}+\frac{377055}{4}c^{2}e^{3}b^{17}
-\frac{1995}{2}e^{36}b^{19}-\frac{12352555215}{2048}e^{24}b^{7}c^{12}-\\
-\frac{10886715}{8}e^{32}b^{15}c^{4})l_{18}(\psi)
+(\frac{1995}{2}e^{36}b^{18}c-\frac{130945815}{256}e^{26}b^{8}c^{11}+\frac{1526175}{8}e^{32}b^{14}c^{5}
+\frac{101846745}{128}e^{28}b^{10}c^{9}-\\
-\frac{101745}{4}e^{34}b^{16}c^{3}-\frac{9258795}{16}e^{30}b^{12}c^{7}
-\frac{72747675}{4096}e^{22}b^{4}c^{15}+\frac{305540235}{2048}e^{24}b^{6}c^{13}-\frac{440895}{131072}c^{19}e^{18}+ \\
+\frac{43648605}{65536}e^{20}b^{2}c^{17})l_{19}(\psi))).\\
\end{gather*}

\[\textbf{Appendix (Item 2). Expression for the Function}K_{14;20}\textbf{:}\]
\allowdisplaybreaks
\begin{gather*}
K_{14;20}(\psi)={}\frac{71645805}{240518168576}(304((\frac{43648605}{128}c^{7}e^{12}b^{12} + 210c^{19}
+\frac{72747675}{16384}c^{3}e^{16}b^{16} - \frac{101846745}{128}c^{9}e^{10}b^{10} + 101745c^{15}e^{4}b^{4} + \\
+\frac{27776385}{32}c^{11}e^{8}b^{8} - \frac{17955}{2}c^{17}e^{2}b^{2} - \frac{3561075}{8}c^{13}e^{6}b^{6}
-\frac{130945815}{2048}c^{5}e^{14}b^{14} - \frac{4849845}{65536}ce^{18}b^{18} )l_{0}(\psi) + (-\frac{101846745}{8}e^{8}b^{7}c^{12} + \\
+\frac{16787925}{4}e^{6}b^{5}c^{14} + \frac{1120314195}{64}c^{10}e^{10}b^{9} + \frac{480134655}{32768}e^{18}b^{17}c^{2}
-\frac{800224425}{2048}e^{16}b^{15}c^{4} + 21945e^{2}bc^{18} + \frac{3360942585}{1024}e^{14}b^{13}c^{6} - \\
-\frac{4849845}{65536}e^{20}b^{19} - \frac{1440403965}{128}c^{8}e^{12}b^{11} - \frac{1119195}{2}e^{4}b^{3}c^{16} )l_{1}(\psi)
+(951615c^{17}e^{4}b^{2} + \frac{24923353455}{2048}e^{16}b^{14}c^{5} -\\
- \frac{129243519405}{2048}e^{14}b^{12}c^{7}
+\frac{567431865}{4}c^{9}e^{12}b^{10} - \frac{17955}{2}c^{19}e^{2} - \frac{18804612645}{128}c^{11}e^{10}b^{8} 
-\frac{112021245}{8}e^{6}b^{4}c^{15} + \\
+\frac{557664345}{8}c^{13}e^{8}b^{6} - \frac{56612240685}{65536}e^{18}b^{16}c^{3} 
+\frac{480134655}{32768}e^{20}b^{18}c)l_{2}(\psi) + (\frac{41410215}{2}c^{16}e^{6}b^{3} + \\
+\frac{92345898645}{4096}c^{4}e^{18}b^{15}
-\frac{377545883715}{2048}e^{16}b^{13}c^{6} + \frac{312567660405}{512}e^{14}b^{11}c^{8} 
-\frac{116032541625}{128}c^{10}e^{12}b^{9} + \frac{9879134265}{16}c^{12}e^{10}b^{7} -\\
-\frac{1473979815}{8}e^{8}b^{5}c^{14}
-\frac{1119195}{2}e^{4}bc^{18} - \frac{56612240685}{65536}e^{20}b^{17}c^{2} + \frac{72747675}{16384}e^{22}b^{19} )l_{3}(\psi)
+(\frac{4057895835}{16}c^{15}e^{8}b^{4} -\\
-\frac{5068345066785}{16384}c^{5}e^{18}b^{14}
+\frac{6356560895685}{4096}e^{16}b^{12}c^{7} - \frac{6844814191215}{2048}e^{14}b^{10}c^{9}
+\frac{92345898645}{4096}c^{3}e^{20}b^{16} +\\
+\frac{418252837275}{128}c^{11}e^{12}b^{8} + 101745c^{19}e^{4}
-\frac{91464786945}{64}c^{13}e^{10}b^{6} - \frac{112021245}{8}e^{6}b^{2}c^{17} - \frac{800224425}{2048}e^{22}b^{18}c)l_{4}(\psi) + \\
+(\frac{60359305545}{32}e^{10}b^{5}c^{14} - \frac{5068345066785}{16384}c^{4}e^{20}b^{15}
+\frac{20061466289865}{8192}c^{6}e^{18}b^{13} - \frac{15947672565825}{2048}e^{16}b^{11}c^{8} + \\
+\frac{24923353455}{2048}e^{22}b^{17}c^{2} + \frac{11235631061655}{1024}e^{14}b^{9}c^{10} 
-\frac{445957797285}{64}c^{12}e^{12}b^{7} - \frac{1473979815}{8}e^{8}b^{3}c^{16} + \frac{16787925}{4}e^{6}bc^{18} - \\
-\frac{130945815}{2048}e^{24}b^{19} )l_{5}(\psi) + (\frac{71662361085}{8}c^{13}e^{12}b^{6}
-\frac{377545883715}{2048}e^{22}b^{16}c^{3} + \frac{20061466289865}{8192}c^{5}e^{20}b^{14} - \\
-\frac{193475919001365}{16384}c^{7}e^{18}b^{12} + \frac{3360942585}{1024}e^{24}b^{18}c
+\frac{49577467764825}{2048}e^{16}b^{10}c^{9} - \frac{45518078720205}{2048}e^{14}b^{8}c^{11} - \\
-\frac{91464786945}{64}c^{15}e^{10}b^{4} + \frac{557664345}{8}c^{17}e^{8}b^{2} - \frac{3561075}{8}c^{19}e^{6} )l_{6}(\psi)
+(\frac{43648605}{128}e^{26}b^{19} + \frac{7193712040515}{256}c^{12}e^{14}b^{7} - \\
-\frac{129243519405}{2048}e^{24}b^{17}c^{2}
+\frac{6356560895685}{4096}e^{22}b^{15}c^{4} - \frac{193475919001365}{16384}c^{6}e^{20}b^{13}
+\frac{146612477846835}{4096}c^{8}e^{18}b^{11} -\\
-\frac{97410839120595}{2048}e^{16}b^{9}c^{10}
-\frac{445957797285}{64}c^{14}e^{12}b^{5} + \frac{9879134265}{16}c^{16}e^{10}b^{3} - \frac{101846745}{8}e^{8}bc^{18} )l_{7}(\psi) + \\
+(\frac{487779442306065}{8192}c^{11}e^{16}b^{8} + \frac{312567660405}{512}e^{24}b^{16}c^{3}
-\frac{15947672565825}{2048}e^{22}b^{14}c^{5} + \frac{146612477846835}{4096}c^{7}e^{20}b^{12} - \\
-\frac{1440403965}{128}e^{26}b^{18}c - \frac{2271939876841635}{32768}c^{9}e^{18}b^{10}
-\frac{45518078720205}{2048}e^{14}b^{6}c^{13} + \frac{418252837275}{128}c^{15}e^{12}b^{4} - \\
-\frac{18804612645}{128}c^{17}e^{10}b^{2} + \frac{27776385}{32}c^{19}e^{8} )l_{8}(\psi)
+(\frac{1415489617707165}{16384}c^{10}e^{18}b^{9} + \frac{567431865}{4}e^{26}b^{17}c^{2} - \\
-\frac{6844814191215}{2048}e^{24}b^{15}c^{4} + \frac{49577467764825}{2048}e^{22}b^{13}c^{6}
-\frac{2271939876841635}{32768}c^{8}e^{20}b^{11} - \frac{97410839120595}{2048}e^{16}b^{7}c^{12} + \\
+\frac{11235631061655}{1024}e^{14}b^{5}c^{14} - \frac{116032541625}{128}c^{16}e^{12}b^{3} 
-\frac{101846745}{128}e^{28}b^{19} + \frac{1120314195}{64}c^{18}e^{10}b)l_{9}(\psi) + \\
+(\frac{1415489617707165}{16384}c^{9}e^{20}b^{10} - \frac{116032541625}{128}e^{26}b^{16}c^{3} 
+\frac{11235631061655}{1024}e^{24}b^{14}c^{5} - \frac{97410839120595}{2048}e^{22}b^{12}c^{7} - \\
-\frac{2271939876841635}{32768}c^{11}e^{18}b^{8} + \frac{49577467764825}{2048}e^{16}b^{6}c^{13} 
-\frac{6844814191215}{2048}e^{14}b^{4}c^{15} + \frac{567431865}{4}c^{17}e^{12}b^{2} - \\
-\frac{101846745}{128}c^{19}e^{10}
+\frac{1120314195}{64}e^{28}b^{18}c)l_{10}(\psi) + (\frac{487779442306065}{8192}c^{8}e^{22}b^{10}
+\frac{418252837275}{128}e^{26}b^{15}c^{4} - \\
-\frac{45518078720205}{2048}e^{24}b^{13}c^{6}
-\frac{2271939876841635}{32768}c^{10}e^{20}b^{9} - \frac{1440403965}{128}c^{18}e^{12}b
+\frac{146612477846835}{4096}c^{12}e^{18}b^{7} - \\
-\frac{15947672565825}{2048}e^{16}b^{5}c^{14} 
+\frac{312567660405}{512}e^{14}b^{3}c^{16} - \frac{18804612645}{128}e^{28}b^{17}c^{2} 
+\frac{27776385}{32}e^{30}b^{19} )l_{11}(\psi) +\\
+(\frac{43648605}{128}c^{19}e^{12} + \frac{7193712040515}{256}c^{7}e^{24}b^{12}
-\frac{445957797285}{64}e^{26}b^{14}c^{5} - \frac{97410839120595}{2048}e^{22}b^{10}c^{9} + \\
+\frac{146612477846835}{4096}c^{11}e^{20}b^{8} - \frac{193475919001365}{16384}c^{13}e^{18}b^{6}
+\frac{6356560895685}{4096}e^{16}b^{4}c^{15} - \frac{129243519405}{2048}e^{14}b^{2}c^{17} + \\
+\frac{9879134265}{16}e^{28}b^{16}c^{3} - \frac{101846745}{8}e^{30}b^{18}c)l_{12}(\psi) + (\frac{71662361085}{8}c^{6}e^{26}b^{13}
-\frac{45518078720205}{2048}e^{24}b^{11}c^{8} + \\
+\frac{49577467764825}{2048}e^{22}b^{9}c^{10}
+\frac{3360942585}{1024}e^{14}bc^{18} - \frac{193475919001365}{16384}c^{12}e^{20}b^{7}
+\frac{20061466289865}{8192}c^{14}e^{18}b^{5} -\\
-\frac{377545883715}{2048}e^{16}b^{3}c^{16} -
\frac{91464786945}{64}e^{28}b^{15}c^{4} + \frac{557664345}{8}c^{2}e^{30}b^{17} - \frac{3561075}{8}e^{32}b^{19} )l_{13}(\psi) + \\
(\frac{60359305545}{32}e^{28}b^{14}c^{5} - \frac{445957797285}{64}e^{26}b^{12}c^{7}
+\frac{11235631061655}{1024}e^{24}b^{10}c^{9} + \frac{24923353455}{2048}e^{16}b^{2}c^{17} - \\
-\frac{15947672565825}{2048}e^{22}b^{8}c^{11} + \frac{20061466289865}{8192}c^{13}e^{20}b^{6} 
-\frac{5068345066785}{16384}c^{15}e^{18}b^{4} - \frac{1473979815}{8}e^{30}b^{16}c^{3} + \frac{16787925}{4}e^{32}b^{18}c - \\
-\frac{130945815}{2048}c^{19}e^{14} )l_{14}(\psi) + (\frac{4057895835}{16}c^{4}e^{30}b^{15} + \frac{418252837275}{128}e^{26}b^{11}c^{8}
+\frac{92345898645}{4096}c^{16}e^{18}b^{3} -\\
-\frac{6844814191215}{2048}e^{24}b^{9}c^{10}
+\frac{6356560895685}{4096}e^{22}b^{7}c^{12} - \frac{5068345066785}{16384}c^{14}e^{20}b^{5}
-\frac{91464786945}{64}e^{28}b^{13}c^{6} - \\
-\frac{112021245}{8}e^{32}b^{17}c^{2} + 101745e^{34}b^{19}
-\frac{800224425}{2048}e^{16}bc^{18} )l_{15}(\psi) + (\frac{41410215}{2}c^{3}e^{32}b^{16} - \frac{116032541625}{128}e^{26}b^{10}c^{9} + \\
+\frac{312567660405}{512}e^{24}b^{8}c^{11} - \frac{377545883715}{2048}e^{22}b^{6}c^{13} 
+\frac{92345898645}{4096}c^{15}e^{20}b^{4} + \frac{9879134265}{16}e^{28}b^{12}c^{7} - \frac{1119195}{2}e^{34}b^{18}c - \\
-\frac{1473979815}{8}e^{30}b^{14}c^{5} - \frac{56612240685}{65536}e^{18}b^{2}c^{17} + \frac{72747675}{16384}c^{19}e^{16} )l_{16}(\psi)
+(951615c^{2}e^{34}b^{17} + \frac{567431865}{4}e^{26}b^{9}c^{10} - \\
-\frac{129243519405}{2048}e^{24}b^{7}c^{12}
+\frac{24923353455}{2048}e^{22}b^{5}c^{14} - \frac{18804612645}{128}e^{28}b^{11}c^{8} - \frac{17955}{2}e^{36}b^{19}
-\frac{112021245}{8}e^{32}b^{15}c^{4} +\\
+\frac{557664345}{8}c^{6}e^{30}b^{13} - \frac{56612240685}{65536}e^{20}b^{3}c^{16} 
+\frac{480134655}{32768}e^{18}bc^{18} )l_{17}(\psi) + (\frac{16787925}{4}e^{32}b^{14}c^{5} - \frac{101846745}{8}e^{30}b^{12}c^{7} + \\
+\frac{3360942585}{1024}e^{24}b^{6}c^{13} - \frac{800224425}{2048}e^{22}b^{4}c^{15} + \frac{1120314195}{64}e^{28}b^{10}c^{9}
+\frac{480134655}{32768}e^{20}b^{2}c^{17} + 21945e^{36}b^{18}c -\\
-\frac{1440403965}{128}e^{26}b^{8}c^{11}
-\frac{4849845}{65536}c^{19}e^{18} - \frac{1119195}{2}e^{34}b^{16}c^{3} )l_{18}(\psi) + (210e^{38}b^{19}
-\frac{101846745}{128}e^{28}b^{9}c^{10} + 101745e^{34}b^{15}c^{4} + \\
+\frac{27776385}{32}e^{30}b^{11}c^{8}
-\frac{17955}{2}e^{36}b^{17}c^{2} - \frac{3561075}{8}e^{32}b^{13}c^{6} - \frac{130945815}{2048}e^{24}b^{5}c^{14}
+\frac{43648605}{128}e^{26}b^{7}c^{12} - \frac{4849845}{65536}e^{20}bc^{18} +\\
+\frac{72747675}{16384}e^{22}b^{3}c^{16} )l_{19}(\psi)
+892i\Omega e((-\frac{305540235}{2048}e^{12}b^{13}c^{6} - \frac{1995}{2}bc^{18} - \frac{43648605}{65536}e^{16}b^{17}c^{2} + \\
+\frac{130945815}{256}e^{10}b^{11}c^{8} - \frac{1526175}{8}e^{4}b^{5}c^{14} - \frac{101846745}{128}e^{8}b^{9}c^{10}
+\frac{101745}{4}e^{2}b^{3}c^{16} + \frac{9258795}{16}e^{6}b^{7}c^{12} + \frac{72747675}{4096}e^{14}b^{15}c^{4} + \\
+\frac{440895}{131072}e^{18}b^{19} )l_{0}(\psi) + (-\frac{107544465}{16}c^{13}e^{6}b^{6} + \frac{1805465025}{128}c^{11}e^{8}b^{8}
-\frac{3477338865}{256}c^{9}e^{10}b^{10} + \frac{5397877485}{65536}c^{3}e^{16}b^{16} - \\
-\frac{182971425}{131072}ce^{18}b^{18} + \frac{1995}{2}c^{19} - \frac{377055}{4}c^{17}e^{2}b^{2} 
-\frac{4757697945}{4096}c^{5}e^{14}b^{14} + \frac{12352555215}{2048}c^{7}e^{12}b^{12} + \frac{10886715}{8}c^{15}e^{4}b^{4} )l_{1}(\psi) + \\
+(-\frac{13125105}{4}e^{4}b^{3}c^{16} + \frac{377055}{4}e^{2}bc^{18} - \frac{94673824245}{1024}c^{8}e^{12}b^{11} 
+\frac{113966507655}{4096}e^{14}b^{13}c^{6} - \frac{27804161385}{8192}e^{16}b^{15}c^{4} + \\
+\frac{35311721445}{256}c^{10}e^{10}b^{9} + \frac{456936795}{16}e^{6}b^{5}c^{14} - \frac{3027625965}{32}e^{8}b^{7}c^{12}
-\frac{43648605}{65536}e^{20}b^{19} + \frac{17011051785}{131072}e^{18}b^{17}c^{2} )l_{2}(\psi) + \\
+(-\frac{910922985}{16}e^{6}b^{4}c^{15} + \frac{13125105}{4}c^{17}e^{4}b^{2} - \frac{1355478329205}{4096}e^{14}b^{12}c^{7} 
+\frac{734504175405}{1024}c^{9}e^{12}b^{10} - \frac{624776873175}{131072}e^{18}b^{16}c^{3} + \\
+\frac{537794462205}{8192}e^{16}b^{14}c^{5} - \frac{181153614915}{256}c^{11}e^{10}b^{8}
+\frac{10038670425}{32}c^{13}e^{8}b^{6} - \frac{101745}{4}c^{19}e^{2} + \frac{5397877485}{65536}e^{20}b^{18}c)l_{3}(\psi) + \\
+(-\frac{36011930535}{64}e^{8}b^{5}c^{14} + \frac{910922985}{16}c^{16}e^{6}b^{3}
-\frac{6457301876025}{2048}c^{10}e^{12}b^{9} + \frac{9067517554095}{4096}e^{14}b^{11}c^{8} - \\
-\frac{11315478909735}{16384}e^{16}b^{13}c^{6} + \frac{2841078440655}{32768}c^{4}e^{18}b^{15}
-\frac{27804161385}{8192}e^{20}b^{17}c^{2} + \frac{260186689035}{128}c^{12}e^{10}b^{7} - \frac{10886715}{8}e^{4}bc^{18} + \\
+\frac{72747675}{4096}e^{22}b^{19} )l_{4}(\psi) + (\frac{36011930535}{64}c^{15}e^{8}b^{4}
-\frac{437433194205}{128}c^{13}e^{10}b^{6} + \frac{16976779693965}{2048}c^{11}e^{12}b^{8} - \\
-\frac{36376092677925}{4096}e^{14}b^{10}c^{9} + \frac{70102234897695}{16384}e^{16}b^{12}c^{7}
-\frac{28796615589105}{32768}c^{5}e^{18}b^{14} + \frac{537794462205}{8192}c^{3}e^{20}b^{16} - \\
-\frac{4757697945}{4096}e^{22}b^{18}c - \frac{456936795}{16}e^{6}b^{2}c^{17} + \frac{1526175}{8}c^{19}e^{4} )l_{5}(\psi)
+(-\frac{6848147357415}{512}c^{12}e^{12}b^{7} + \frac{437433194205}{128}e^{10}b^{5}c^{14} - \\
-\frac{305540235}{2048}e^{24}b^{19} + \frac{90588503960955}{4096}e^{14}b^{9}c^{10} 
-\frac{133892925160995}{8192}e^{16}b^{11}c^{8} + \frac{174186589920525}{32768}c^{6}e^{18}b^{13}-\\
-\frac{11315478909735}{16384}c^{4}e^{20}b^{15} + \frac{113966507655}{4096}e^{22}b^{17}c^{2}
-\frac{10038670425}{32}e^{8}b^{3}c^{16} + \frac{107544465}{16}e^{6}bc^{18} )l_{6}(\psi) + \\
+(-\frac{142481264361345}{4096}e^{14}b^{8}c^{11} + \frac{6848147357415}{512}c^{13}e^{12}b^{6} 
+\frac{323723094429735}{8192}e^{16}b^{10}c^{9} - \frac{655162385283315}{32768}c^{7}e^{18}b^{12} + \\
+\frac{70102234897695}{16384}c^{5}e^{20}b^{14} - \frac{1355478329205}{4096}e^{22}b^{16}c^{3}
+\frac{12352555215}{2048}e^{24}b^{18}c - \frac{260186689035}{128}c^{15}e^{10}b^{4} - \frac{9258795}{16}c^{19}e^{6} + \\
+\frac{3027625965}{32}c^{17}e^{8}b^{2} )l_{7}(\psi) + (-\frac{2008258828631055}{32768}e^{16}b^{9}c^{10} 
+\frac{142481264361345}{4096}c^{12}e^{14}b^{7} - \frac{16976779693965}{2048}c^{14}e^{12}b^{5} + \\
+\frac{3138386978624625}{65536}c^{8}e^{18}b^{11} - \frac{133892925160995}{8192}c^{6}e^{20}b^{13} 
+\frac{9067517554095}{4096}e^{22}b^{15}c^{4} - \frac{94673824245}{1024}e^{24}b^{17}c^{2} + \\
+\frac{181153614915}{256}c^{16}e^{10}b^{3} + \frac{130945815}{256}e^{26}b^{19} - \frac{1805465025}{128}e^{8}bc^{18} )l_{8}(\psi)
+(-\frac{4845557363581935}{65536}c^{9}e^{18}b^{10} + \\
+\frac{2008258828631055}{32768}c^{11}e^{16}b^{8} 
+\frac{6457301876025}{2048}c^{15}e^{12}b^{4} - \frac{90588503960955}{4096}e^{14}b^{6}c^{13} 
+\frac{323723094429735}{8192}c^{7}e^{20}b^{12} -\\
-\frac{36376092677925}{4096}e^{22}b^{14}c^{5}
+\frac{734504175405}{1024}e^{24}b^{16}c^{3} - \frac{35311721445}{256}c^{17}e^{10}b^{2}
-\frac{3477338865}{256}e^{26}b^{18}c + \frac{101846745}{128}c^{19}e^{8} )l_{9}(\psi) + \\
+(-\frac{2008258828631055}{32768}c^{8}e^{20}b^{11} + \frac{4845557363581935}{65536}c^{10}e^{18}b^{9}
-\frac{734504175405}{1024}c^{1}e^{12}b^{3} + \frac{36376092677925}{4096}e^{14}b^{5}c^{14} - \\
-\frac{323723094429735}{8192}e^{16}b^{7}c^{12} + \frac{90588503960955}{4096}e^{22}b^{13}c^{6}
-\frac{6457301876025}{2048}e^{24}b^{15}c^{4} + \frac{3477338865}{256}c^{18}e^{10}b + \\
+\frac{35311721445}{256}e^{26}b^{17}c^{2} - \frac{101846745}{128}e^{28}b^{19} )l_{10}(\psi)
+(-\frac{142481264361345}{4096}e^{22}b^{12}c^{7} + \frac{2008258828631055}{32768}c^{9}e^{20}b^{10} + \\
+\frac{94673824245}{1024}c^{17}e^{12}b^{2} - \frac{9067517554095}{4096}e^{14}b^{4}c^{15}
+\frac{133892925160995}{8192}e^{16}b^{6}c^{13} - \frac{3138386978624625}{65536}c^{11}e^{18}b^{8} + \\
+\frac{16976779693965}{2048}e^{24}b^{14}c^{5} - \frac{181153614915}{256}e^{26}b^{16}c^{3}
-\frac{130945815}{256}c^{19}e^{10} + \frac{1805465025}{128}e^{28}b^{18}c)l_{11}(\psi) + \\
+(-\frac{6848147357415}{512}e^{24}b^{13}c^{6} + \frac{142481264361345}{4096}c^{8}e^{22}b^{11}
-\frac{12352555215}{2048}c^{18}e^{12}b + \frac{1355478329205}{4096}e^{14}b^{3}c^{16} - \\
-\frac{70102234897695}{16384}e^{16}b^{5}c^{14} + \frac{655162385283315}{32768}c^{12}e^{18}b^{7}
-\frac{323723094429735}{8192}c^{10}e^{20}b^{9} + \frac{260186689035}{128}e^{26}b^{15}c^{4} + \\
+\frac{9258795}{16}e^{30}b^{19} - \frac{3027625965}{32}e^{28}b^{17}c^{2} )l_{12}(\psi)
+(-\frac{437433194205}{128}e^{26}b^{14}c^{5} + \frac{6848147357415}{512}c^{7}e^{24}b^{12} + \\
+\frac{305540235}{2048}c^{19}e^{12} - \frac{113966507655}{4096}e^{14}b^{2}c^{17}
+\frac{11315478909735}{16384}e^{16}b^{4}c^{15} - \frac{174186589920525}{32768}c^{13}e^{18}b^{6} + \\
+\frac{133892925160995}{8192}c^{11}e^{20}b^{8} - \frac{90588503960955}{4096}e^{22}b^{10}c^{9} 
+\frac{10038670425}{32}e^{28}b^{16}c^{3} - \frac{107544465}{16}e^{30}b^{18}c)l_{13}(\psi) + \\
+(\frac{437433194205}{128}c^{6}e^{26}b^{13} - \frac{36011930535}{64}e^{28}b^{15}c^{4}
+\frac{4757697945}{4096}e^{14}bc^{18} - \frac{537794462205}{8192}e^{16}b^{3}c^{16} + \\
+\frac{28796615589105}{32768}c^{14}e^{18}b^{5} - \frac{70102234897695}{16384}c^{12}e^{20}b^{7}
+\frac{36376092677925}{4096}e^{22}b^{9}c^{10} - \frac{16976779693965}{2048}e^{24}b^{11}c^{8} + \\
+\frac{456936795}{16}c^{2}e^{30}b^{17} - \frac{1526175}{8}e^{32}b^{19} )l_{14}(\psi) + (-\frac{910922985}{16}e^{30}b^{16}c^{3}
+\frac{36011930535}{64}e^{28}b^{14}c^{5} + \frac{27804161385}{8192}e^{16}b^{2}c^{17} - \\
-\frac{2841078440655}{32768}c^{15}e^{18}b^{4} + \frac{11315478909735}{16384}c^{13}e^{20}b^{6} 
-\frac{9067517554095}{4096}e^{22}b^{8}c^{11} + \frac{6457301876025}{2048}e^{24}b^{10}c^{9} - \\
-\frac{260186689035}{128}e^{26}b^{12}c^{7} + \frac{10886715}{8}e^{32}b^{18}c - \frac{72747675}{4096}c^{19}e^{14} )l_{15}(\psi)
+(-\frac{13125105}{4}e^{32}b^{17}c^{2} + \frac{910922985}{16}c^{4}e^{30}b^{15} -\\
-\frac{537794462205}{8192}c^{14}e^{20}b^{5}
+\frac{624776873175}{131072}c^{16}e^{18}b^{3} - \frac{734504175405}{1024}e^{24}b^{9}c^{10}
+\frac{1355478329205}{4096}e^{22}b^{7}c^{12} +\\
+\frac{181153614915}{256}e^{26}b^{11}c^{8}
-\frac{10038670425}{32}e^{28}b^{13}c^{6} + \frac{101745}{4}e^{34}b^{19} - \frac{5397877485}{65536}e^{16}bc^{18} )l_{16}(\psi) + \\
+(-\frac{377055}{4}e^{34}b^{18}c + \frac{13125105}{4}c^{3}e^{32}b^{16} + \frac{3027625965}{32}e^{28}b^{12}c^{7}
+\frac{27804161385}{8192}c^{15}e^{20}b^{4} - \frac{113966507655}{4096}e^{22}b^{6}c^{13} + \\
+\frac{94673824245}{1024}e^{24}b^{8}c^{11} - \frac{35311721445}{256}e^{26}b^{10}c^{9}
-\frac{456936795}{16}e^{30}b^{14}c^{5} + \frac{43648605}{65536}c^{19}e^{16} - \frac{17011051785}{131072}e^{18}b^{2}c^{17} )l_{17}(\psi) + \\
+(-\frac{1805465025}{128}e^{28}b^{11}c^{8} + \frac{107544465}{16}c^{6}e^{30}b^{13} - \frac{5397877485}{65536}e^{20}b^{3}c^{16}
+\frac{4757697945}{4096}e^{22}b^{5}c^{14} + \frac{3477338865}{256}e^{26}b^{9}c^{10} +\\
+ \frac{182971425}{131072}e^{18}bc^{18}
+\frac{377055}{4}c^{2}e^{3}b^{17} - \frac{1995}{2}e^{36}b^{19} - \frac{12352555215}{2048}e^{24}b^{7}c^{12} 
-\frac{10886715}{8}e^{32}b^{15}c^{4} )l_{18}(\psi) + (\frac{1995}{2}e^{36}b^{18}c - \\
-\frac{130945815}{256}e^{26}b^{8}c^{11}
+\frac{1526175}{8}e^{32}b^{14}c^{5} + \frac{101846745}{128}e^{28}b^{10}c^{9} - \frac{101745}{4}e^{34}b^{16}c^{3} 
-\frac{9258795}{16}e^{30}b^{12}c^{7} - \\
-\frac{72747675}{4096}e^{22}b^{4}c^{15} + \frac{305540235}{2048}e^{24}b^{6}c^{13}
-\frac{440895}{131072}c^{19}e^{18} + \frac{43648605}{65536}e^{20}b^{2}c^{17} )l_{19}(\psi))).
\end{gather*}

\[\textbf{Appendix (Item 3). Expression for the Function }K_{16;20}\textbf{:}\]
\allowdisplaybreaks
\begin{gather*}
K_{16;20}(\psi)={}\frac{3411705}{34359738368}(779((\frac{43648605}{128}c^{7}e^{12}b^{12} + 210c^{19} +
\frac{72747675}{16384}c^{3}e^{16}b^{16} - \frac{101846745}{128}c^{9}e^{10}b^{10} + \\ 
+101745c^{15}e^{4}b^{4} +
+\frac{27776385}{32}c^{11}e^{8}b^{8} - \frac{17955}{2}c^{17}e^{2}b^{2} - \frac{3561075}{8}c^{13}e^{6}b^{6}
-\frac{130945815}{2048}c^{5}e^{14}b^{14} - \frac{4849845}{65536}ce^{18}b^{18} )l_{0}(\psi) +\\
+(-\frac{101846745}{8}e^{8}b^{7}c^{12} 
+\frac{16787925}{4}e^{6}b^{5}c^{14} + \frac{1120314195}{64}c^{10}e^{10}b^{9} + \frac{480134655}{32768}e^{18}b^{17}c^{2} 
-\frac{800224425}{2048}e^{16}b^{15}c^{4} +\\ +21945e^{2}bc^{18} + \frac{3360942585}{1024}e^{14}b^{13}c^{6}
-\frac{4849845}{65536}e^{20}b^{19} - \frac{1440403965}{128}c^{8}e^{12}b^{11} - \frac{1119195}{2}e^{4}b^{3}c^{16} )l_{1}(\psi) + \\
+(951615c^{17}e^{4}b^{2} + \frac{24923353455}{2048}e^{16}b^{14}c^{5} - \frac{129243519405}{2048}e^{14}b^{12}c^{7} + \frac{567431865}{4}c^{9}e^{12}b^{10} - \frac{17955}{2}c^{19}e^{2} - \\
-\frac{18804612645}{128}c^{11}e^{10}b^{8} 
-\frac{112021245}{8}e^{6}b^{4}c^{15} + \frac{557664345}{8}c^{13}e^{8}b^{6} - \frac{56612240685}{65536}e^{18}b^{16}c^{3} 
+\frac{480134655}{32768}e^{20}b^{18}c)l_{2}(\psi) +\\ +(\frac{41410215}{2}c^{16}e^{6}b^{3} + \frac{92345898645}{4096}c^{4}e^{18}b^{15} 
-\frac{377545883715}{2048}e^{16}b^{13}c^{6} + \frac{312567660405}{512}e^{14}b^{11}c^{8} 
-\frac{116032541625}{128}c^{10}e^{12}b^{9} + \\
+\frac{9879134265}{16}c^{12}e^{10}b^{7} - \frac{1473979815}{8}e^{8}b^{5}c^{14}
-\frac{1119195}{2}e^{4}bc^{18} - \frac{56612240685}{65536}e^{20}b^{17}c^{2} + \frac{72747675}{16384}e^{22}b^{19} )l_{3}(\psi) + \\
+(\frac{4057895835}{16}c^{15}e^{8}b^{4} - \frac{5068345066785}{16384}c^{5}e^{18}b^{14} 
+\frac{6356560895685}{4096}e^{16}b^{12}c^{7} - \frac{6844814191215}{2048}e^{14}b^{10}c^{9} + \\
+\frac{92345898645}{4096}c^{3}e^{20}b^{16} + \frac{418252837275}{128}c^{11}e^{12}b^{8} + 101745c^{19}e^{4}
-\frac{91464786945}{64}c^{13}e^{10}b^{6} - \frac{112021245}{8}e^{6}b^{2}c^{17} - \\
-\frac{800224425}{2048}e^{22}b^{18}c)l_{4}(\psi) 
+(\frac{60359305545}{32}e^{10}b^{5}c^{14} - \frac{5068345066785}{16384}c^{4}e^{20}b^{15} 
+\frac{20061466289865}{8192}c^{6}e^{18}b^{13} - \\
-\frac{15947672565825}{2048}e^{16}b^{11}c^{8} 
+\frac{24923353455}{2048}e^{22}b^{17}c^{2} + \frac{11235631061655}{1024}e^{14}b^{9}c^{10}
-\frac{445957797285}{64}c^{12}e^{12}b^{7} - \\
-\frac{1473979815}{8}e^{8}b^{3}c^{16} + \frac{16787925}{4}e^{6}bc^{18}
-\frac{130945815}{2048}e^{24}b^{19} )l_{5}(\psi) + (\frac{71662361085}{8}c^{13}e^{12}b^{6} - \frac{377545883715}{2048}e^{22}b^{16}c^{3} + \\
+\frac{20061466289865}{8192}c^{5}e^{20}b^{14} - \frac{193475919001365}{16384}c^{7}e^{18}b^{12} +\frac{3360942585}{1024}e^{24}b^{18}c + \frac{49577467764825}{2048}e^{16}b^{10}c^{9} - \\
-\frac{45518078720205}{2048}e^{14}b^{8}c^{11} - \frac{91464786945}{64}c^{15}e^{10}b^{4} 
+\frac{557664345}{8}c^{17}e^{8}b^{2} - \frac{3561075}{8}c^{19}e^{6} )l_{6}(\psi) + (\frac{43648605}{128}e^{26}b^{19} + \\
+\frac{7193712040515}{256}c^{12}e^{14}b^{7} - \frac{129243519405}{2048}e^{24}b^{17}c^{2} 
+\frac{6356560895685}{4096}e^{22}b^{15}c^{4} - \frac{193475919001365}{16384}c^{6}e^{20}b^{13} + \\
+\frac{146612477846835}{4096}c^{8}e^{18}b^{11} - \frac{97410839120595}{2048}e^{16}b^{9}c^{10} 
-\frac{445957797285}{64}c^{14}e^{12}b^{5} + \frac{9879134265}{16}c^{16}e^{10}b^{3} - \\
-\frac{101846745}{8}e^{8}bc^{18} )l_{7}(\psi) 
+(\frac{487779442306065}{8192}c^{11}e^{16}b^{8} + \frac{312567660405}{512}e^{24}b^{16}c^{3} 
-\frac{15947672565825}{2048}e^{22}b^{14}c^{5} +\\
+\frac{146612477846835}{4096}c^{7}e^{20}b^{12} 
-\frac{1440403965}{128}e^{26}b^{18}c - \frac{2271939876841635}{32768}c^{9}e^{18}b^{10} 
-\frac{45518078720205}{2048}e^{14}b^{6}c^{13} +\\
+\frac{418252837275}{128}c^{15}e^{12}b^{4} 
-\frac{18804612645}{128}c^{17}e^{10}b^{2} + \frac{27776385}{32}c^{19}e^{8} )l_{8}(\psi)
+(\frac{1415489617707165}{16384}c^{10}e^{18}b^{9} +\\
+\frac{567431865}{4}e^{26}b^{17}c^{2}
-\frac{6844814191215}{2048}e^{24}b^{15}c^{4} + \frac{49577467764825}{2048}e^{22}b^{13}c^{6} 
-\frac{2271939876841635}{32768}c^{8}e^{20}b^{11} - \\\
-\frac{97410839120595}{2048}e^{16}b^{7}c^{12} 
+\frac{11235631061655}{1024}e^{14}b^{5}c^{14} - \frac{116032541625}{128}c^{16}e^{12}b^{3}
-\frac{101846745}{128}e^{28}b^{19} + \\
+\frac{1120314195}{64}c^{18}e^{10}b)l_{9}(\psi)
+(\frac{1415489617707165}{16384}c^{9}e^{20}b^{10} - \frac{116032541625}{128}e^{26}b^{16}c^{3} 
+\frac{11235631061655}{1024}e^{24}b^{14}c^{5} -\\
-\frac{97410839120595}{2048}e^{22}b^{12}c^{7}
-\frac{2271939876841635}{32768}c^{11}e^{18}b^{8} + \frac{49577467764825}{2048}e^{16}b^{6}c^{13} 
-\frac{6844814191215}{2048}e^{14}b^{4}c^{15} + \\
+\frac{567431865}{4}c^{17}e^{12}b^{2} - \frac{101846745}{128}c^{19}e^{10} 
+\frac{1120314195}{64}e^{28}b^{18}c)l_{10}(\psi) + (\frac{487779442306065}{8192}c^{8}e^{22}b^{10} + \\
+\frac{418252837275}{128}e^{26}b^{15}c^{4} - \frac{45518078720205}{2048}e^{24}b^{13}c^{6}
-\frac{2271939876841635}{32768}c^{10}e^{20}b^{9} - \frac{1440403965}{128}c^{18}e^{12}b + \\
+\frac{146612477846835}{4096}c^{12}e^{18}b^{7} - \frac{15947672565825}{2048}e^{16}b^{5}c^{14}
+\frac{312567660405}{512}e^{14}b^{3}c^{16} - \frac{18804612645}{128}e^{28}b^{17}c^{2} + \\
+\frac{27776385}{32}e^{30}b^{19} )l_{11}(\psi) 
+(\frac{43648605}{128}c^{19}e^{12} + \frac{7193712040515}{256}c^{7}e^{24}b^{12} - \frac{445957797285}{64}e^{26}b^{14}c^{5} - \\
-\frac{97410839120595}{2048}e^{22}b^{10}c^{9} + \frac{146612477846835}{4096}c^{11}e^{20}b^{8} 
-\frac{193475919001365}{16384}c^{13}e^{18}b^{6} + \frac{6356560895685}{4096}e^{16}b^{4}c^{15} - \\
-\frac{129243519405}{2048}e^{14}b^{2}c^{17} + \frac{9879134265}{16}e^{28}b^{16}c^{3} -\frac{101846745}{8}e^{30}b^{18}c)l_{12}(\psi) 
+(\frac{71662361085}{8}c^{6}e^{26}b^{13} -\\
-\frac{45518078720205}{2048}e^{24}b^{11}c^{8} 
+\frac{49577467764825}{2048}e^{22}b^{9}c^{10} + \frac{3360942585}{1024}e^{14}bc^{18} 
-\frac{193475919001365}{16384}c^{12}e^{20}b^{7} +\\
+\frac{20061466289865}{8192}c^{14}e^{18}b^{5} 
-\frac{377545883715}{2048}e^{16}b^{3}c^{16} - \frac{91464786945}{64}e^{28}b^{15}c^{4} + \frac{557664345}{8}c^{2}e^{30}b^{17} - \\
-\frac{3561075}{8}e^{32}b^{19} )l_{13}(\psi) + (\frac{60359305545}{32}e^{28}b^{14}c^{5} - \frac{445957797285}{64}e^{26}b^{12}c^{7}
+\frac{11235631061655}{1024}e^{24}b^{10}c^{9} +\\
+\frac{24923353455}{2048}e^{16}b^{2}c^{17} 
-\frac{15947672565825}{2048}e^{22}b^{8}c^{11} + \frac{20061466289865}{8192}c^{13}e^{20}b^{6} 
-\frac{5068345066785}{16384}c^{15}e^{18}b^{4} - \\
-\frac{1473979815}{8}e^{30}b^{16}c^{3} + \frac{16787925}{4}e^{32}b^{18}c  
-\frac{130945815}{2048}c^{19}e^{14} )l_{14}(\psi) 
+(\frac{4057895835}{16}c^{4}e^{30}b^{15} + \frac{418252837275}{128}e^{26}b^{11}c^{8}+\\ 
+\frac{92345898645}{4096}c^{16}e^{18}b^{3} - \frac{6844814191215}{2048}e^{24}b^{9}c^{10} 
+\frac{6356560895685}{4096}e^{22}b^{7}c^{12} - \frac{5068345066785}{16384}c^{14}e^{20}b^{5} - \\
-\frac{91464786945}{64}e^{28}b^{13}c^{6} - \frac{112021245}{8}e^{32}b^{17}c^{2} + 101745e^{34}b^{19} 
-\frac{800224425}{2048}e^{16}bc^{18} )l_{15}(\psi) +\\
+(\frac{41410215}{2}c^{3}e^{32}b^{16} -\frac{116032541625}{128}e^{26}b^{10}c^{9} 
+\frac{312567660405}{512}e^{24}b^{8}c^{11} - \frac{377545883715}{2048}e^{22}b^{6}c^{13} + \\
+\frac{92345898645}{4096}c^{15}e^{20}b^{4} + \frac{9879134265}{16}e^{28}b^{12}c^{7} - \frac{1119195}{2}e^{34}b^{18}c 
-\frac{1473979815}{8}e^{30}b^{14}c^{5} - \frac{56612240685}{65536}e^{18}b^{2}c^{17} + \\
+\frac{72747675}{16384}c^{19}e^{16} )l_{16}(\psi) 
+(951615c^{2}e^{34}b^{17} + \frac{567431865}{4}e^{26}b^{9}c^{10} -\frac{129243519405}{2048}e^{24}b^{7}c^{12} + \\
+\frac{24923353455}{2048}e^{22}b^{5}c^{14} - \frac{18804612645}{128}e^{28}b^{11}c^{8} - \frac{17955}{2}e^{36}b^{19} 
-\frac{112021245}{8}e^{32}b^{15}c^{4} + \frac{557664345}{8}c^{6}e^{30}b^{13} -\\
-\frac{56612240685}{65536}e^{20}b^{3}c^{16} 
+\frac{480134655}{32768}e^{18}bc^{18} )l_{17}(\psi) + (\frac{16787925}{4}e^{32}b^{14}c^{5} - \frac{101846745}{8}e^{30}b^{12}c^{7} +\\
+\frac{3360942585}{1024}e^{24}b^{6}c^{13} - \frac{800224425}{2048}e^{22}b^{4}c^{15} + \frac{1120314195}{64}e^{28}b^{10}c^{9} 
+\frac{480134655}{32768}e^{20}b^{2}c^{17} + 21945e^{36}b^{18}c -\\ 
-\frac{1440403965}{128}e^{26}b^{8}c^{11}
-\frac{4849845}{65536}c^{19}e^{18} -\frac{1119195}{2}e^{34}b^{16}c^{3})l_{18}(\psi) + (210e^{38}b^{19} 
-\frac{101846745}{128}e^{28}b^{9}c^{10} + 101745e^{34}b^{15}c^{4} + \\
+\frac{27776385}{32}e^{30}b^{11}c^{8} 
-\frac{17955}{2}e^{36}b^{17}c^{2} - \frac{3561075}{8}e^{32}b^{13}c^{6} -\frac{130945815}{2048}e^{24}b^{5}c^{14} + \\
\frac{43648605}{128}e^{26}b^{7}c^{12} - \frac{4849845}{65536}e^{20}bc^{18} + \frac{72747675}{16384}e^{22}b^{3}c^{16} )l_{19}(\psi) + \\
+2608i\Omega e((-\frac{305540235}{2048}e^{12}b^{13}c^{6} - \frac{1995}{2}bc^{18} - \frac{43648605}{65536}e^{16}b^{17}c^{2} 
+\frac{130945815}{256}e^{10}b^{11}c^{8} - \frac{1526175}{8}e^{4}b^{5}c^{14} - \\
-\frac{101846745}{128}e^{8}b^{9}c^{10} 
+\frac{101745}{4}e^{2}b^{3}c^{16} + \frac{9258795}{16}e^{6}b^{7}c^{12} + \frac{72747675}{4096}e^{14}b^{15}c^{4} 
+\frac{440895}{131072}e^{18}b^{19} )l_{0}(\psi) +
\\+ (-\frac{107544465}{16}c^{13}e^{6}b^{6} + \frac{1805465025}{128}c^{11}e^{8}b^{8} 
-\frac{3477338865}{256}c^{9}e^{10}b^{10} + \frac{5397877485}{65536}c^{3}e^{16}b^{16} - \\
-\frac{182971425}{131072}ce^{18}b^{18} + \frac{1995}{2}c^{19} - \frac{377055}{4}c^{17}e^{2}b^{2}
-\frac{4757697945}{4096}c^{5}e^{14}b^{14} + \frac{12352555215}{2048}c^{7}e^{12}b^{12} + \frac{10886715}{8}c^{15}e^{4}b^{4} )l_{1}(\psi) + \\
+(-\frac{13125105}{4}e^{4}b^{3}c^{16} + \frac{377055}{4}e^{2}bc^{18} - \frac{94673824245}{1024}c^{8}e^{12}b^{11} 
+\frac{113966507655}{4096}e^{14}b^{13}c^{6} - \frac{27804161385}{8192}e^{16}b^{15}c^{4} + \\
+\frac{35311721445}{256}c^{10}e^{10}b^{9} + \frac{456936795}{16}e^{6}b^{5}c^{14} - \frac{3027625965}{32}e^{8}b^{7}c^{12} 
-\frac{43648605}{65536}e^{20}b^{19} + \frac{17011051785}{131072}e^{18}b^{17}c^{2} )l_{2}(\psi) + \\
+(-\frac{910922985}{16}e^{6}b^{4}c^{15} + \frac{13125105}{4}c^{17}e^{4}b^{2} - \frac{1355478329205}{4096}e^{14}b^{12}c^{7} 
+\frac{734504175405}{1024}c^{9}e^{12}b^{10} - \frac{624776873175}{131072}e^{18}b^{16}c^{3} + \\
+\frac{537794462205}{8192}e^{16}b^{14}c^{5} - \frac{181153614915}{256}c^{11}e^{10}b^{8} 
+\frac{10038670425}{32}c^{13}e^{8}b^{6} - \frac{101745}{4}c^{19}e^{2} + \frac{5397877485}{65536}e^{20}b^{18}c)l_{3}(\psi) + \\
+(-\frac{36011930535}{64}e^{8}b^{5}c^{14} + \frac{910922985}{16}c^{16}e^{6}b^{3} 
-\frac{6457301876025}{2048}c^{10}e^{12}b^{9} + \frac{9067517554095}{4096}e^{14}b^{11}c^{8} - \\
-\frac{11315478909735}{16384}e^{16}b^{13}c^{6} + \frac{2841078440655}{32768}c^{4}e^{18}b^{15} 
-\frac{27804161385}{8192}e^{20}b^{17}c^{2} + \frac{260186689035}{128}c^{12}e^{10}b^{7} - \frac{10886715}{8}e^{4}bc^{18} + \\
+\frac{72747675}{4096}e^{22}b^{19} )l_{4}(\psi) + (\frac{36011930535}{64}c^{15}e^{8}b^{4} 
-\frac{437433194205}{128}c^{13}e^{10}b^{6} + \frac{16976779693965}{2048}c^{11}e^{12}b^{8} - \\
-\frac{36376092677925}{4096}e^{14}b^{10}c^{9} + \frac{70102234897695}{16384}e^{16}b^{12}c^{7} 
-\frac{28796615589105}{32768}c^{5}e^{18}b^{14} + \frac{537794462205}{8192}c^{3}e^{20}b^{16} - \\
-\frac{4757697945}{4096}e^{22}b^{18}c - \frac{456936795}{16}e^{6}b^{2}c^{17} + \frac{1526175}{8}c^{19}e^{4} )l_{5}(\psi) 
+(-\frac{6848147357415}{512}c^{12}e^{12}b^{7} + \frac{437433194205}{128}e^{10}b^{5}c^{14} - \\
-\frac{305540235}{2048}e^{24}b^{19} + \frac{90588503960955}{4096}e^{14}b^{9}c^{10} 
-\frac{133892925160995}{8192}e^{16}b^{11}c^{8} + \frac{174186589920525}{32768}c^{6}e^{18}b^{13} - \\
-\frac{11315478909735}{16384}c^{4}e^{20}b^{15} + \frac{113966507655}{4096}e^{22}b^{17}c^{2} 
-\frac{10038670425}{32}e^{8}b^{3}c^{16} + \frac{107544465}{16}e^{6}bc^{18} )l_{6}(\psi) + \\
+(-\frac{142481264361345}{4096}e^{14}b^{8}c^{11} + \frac{6848147357415}{512}c^{13}e^{12}b^{6} 
+\frac{323723094429735}{8192}e^{16}b^{10}c^{9} - \frac{655162385283315}{32768}c^{7}e^{18}b^{12} + \\
+\frac{70102234897695}{16384}c^{5}e^{20}b^{14} -\frac{1355478329205}{4096}e^{22}b^{16}c^{3} 
+\frac{12352555215}{2048}e^{24}b^{18}c - \frac{260186689035}{128}c^{15}e^{10}b^{4} - \frac{9258795}{16}c^{19}e^{6} + \\
+\frac{3027625965}{32}c^{17}e^{8}b^{2} )l_{7}(\psi) + (-\frac{2008258828631055}{32768}e^{16}b^{9}c^{10} 
+\frac{142481264361345}{4096}c^{12}e^{14}b^{7} -\frac{16976779693965}{2048}c^{14}e^{12}b^{5} + \\
+\frac{3138386978624625}{65536}c^{8}e^{18}b^{11} -\frac{133892925160995}{8192}c^{6}e^{20}b^{13} 
+\frac{9067517554095}{4096}e^{22}b^{15}c^{4} - \frac{94673824245}{1024}e^{24}b^{17}c^{2} + \\
+\frac{181153614915}{256}c^{16}e^{10}b^{3} + \frac{130945815}{256}e^{26}b^{19} -\frac{1805465025}{128}e^{8}bc^{18})l_{8}(\psi) 
+(-\frac{4845557363581935}{65536}c^{9}e^{18}b^{10} + \\
+\frac{2008258828631055}{32768}c^{11}e^{16}b^{8} 
+\frac{6457301876025}{2048}c^{15}e^{12}b^{4} - \frac{90588503960955}{4096}e^{14}b^{6}c^{13} 
+\frac{323723094429735}{8192}c^{7}e^{20}b^{12} -\\
- \frac{36376092677925}{4096}e^{22}b^{14}c^{5} 
+\frac{734504175405}{1024}e^{24}b^{16}c^{3} - \frac{35311721445}{256}c^{17}e^{10}b^{2} 
-\frac{3477338865}{256}e^{26}b^{18}c + \frac{101846745}{128}c^{19}e^{8})l_{9}(\psi) + \\
+(-\frac{2008258828631055}{32768}c^{8}e^{20}b^{11} + \frac{4845557363581935}{65536}c^{10}e^{18}b^{9} 
-\frac{734504175405}{1024}c^{1}e^{12}b^{3} + \frac{36376092677925}{4096}e^{14}b^{5}c^{14} - \\
-\frac{323723094429735}{8192}e^{16}b^{7}c^{12} + \frac{90588503960955}{4096}e^{22}b^{13}c^{6} 
-\frac{6457301876025}{2048}e^{24}b^{15}c^{4} + \frac{3477338865}{256}c^{18}e^{10}b + \\
+\frac{35311721445}{256}e^{26}b^{17}c^{2} -\frac{101846745}{128}e^{28}b^{19})l_{10}(\psi) 
+(-\frac{142481264361345}{4096}e^{22}b^{12}c^{7} + \frac{2008258828631055}{32768}c^{9}e^{20}b^{10} + \\
+\frac{94673824245}{1024}c^{17}e^{12}b^{2} - \frac{9067517554095}{4096}e^{14}b^{4}c^{15} 
+\frac{133892925160995}{8192}e^{16}b^{6}c^{13} - \frac{3138386978624625}{65536}c^{11}e^{18}b^{8} + \\
+\frac{16976779693965}{2048}e^{24}b^{14}c^{5} -\frac{181153614915}{256}e^{26}b^{16}c^{3} 
-\frac{130945815}{256}c^{19}e^{10} + \frac{1805465025}{128}e^{28}b^{18}c)l_{11}(\psi) + \\
+(-\frac{6848147357415}{512}e^{24}b^{13}c^{6} + \frac{142481264361345}{4096}c^{8}e^{22}b^{11} 
-\frac{12352555215}{2048}c^{18}e^{12}b + \frac{1355478329205}{4096}e^{14}b^{3}c^{16} - \\
-\frac{70102234897695}{16384}e^{16}b^{5}c^{14} + \frac{655162385283315}{32768}c^{12}e^{18}b^{7} 
-\frac{323723094429735}{8192}c^{10}e^{20}b^{9} + \frac{260186689035}{128}e^{26}b^{15}c^{4} + \\
+\frac{9258795}{16}e^{30}b^{19} - \frac{3027625965}{32}e^{28}b^{17}c^{2})l_{12}(\psi) 
+(-\frac{437433194205}{128}e^{26}b^{14}c^{5} + \frac{6848147357415}{512}c^{7}e^{24}b^{12} + \\
+\frac{305540235}{2048}c^{19}e^{12} - \frac{113966507655}{4096}e^{14}b^{2}c^{17} 
+\frac{11315478909735}{16384}e^{16}b^{4}c^{15} - \frac{174186589920525}{32768}c^{13}e^{18}b^{6} + \\
+\frac{133892925160995}{8192}c^{11}e^{20}b^{8} - \frac{90588503960955}{4096}e^{22}b^{10}c^{9} 
+\frac{10038670425}{32}e^{28}b^{16}c^{3} - \frac{107544465}{16}e^{30}b^{18}c)l_{13}(\psi) + \\
+(\frac{437433194205}{128}c^{6}e^{26}b^{13} - \frac{36011930535}{64}e^{28}b^{15}c^{4} 
+\frac{4757697945}{4096}e^{14}bc^{18} - \frac{537794462205}{8192}e^{16}b^{3}c^{16} + \\
+\frac{28796615589105}{32768}c^{14}e^{18}b^{5} - \frac{70102234897695}{16384}c^{12}e^{20}b^{7} 
+\frac{36376092677925}{4096}e^{22}b^{9}c^{10} - \frac{16976779693965}{2048}e^{24}b^{11}c^{8} + \\
+\frac{456936795}{16}c^{2}e^{30}b^{17} - \frac{1526175}{8}e^{32}b^{19} )l_{14}(\psi) + (-\frac{910922985}{16}e^{30}b^{16}c^{3} 
+\frac{36011930535}{64}e^{28}b^{14}c^{5} + \frac{27804161385}{8192}e^{16}b^{2}c^{17} - \\
-\frac{2841078440655}{32768}c^{15}e^{18}b^{4} + \frac{11315478909735}{16384}c^{13}e^{20}b^{6} 
-\frac{9067517554095}{4096}e^{22}b^{8}c^{11} + \frac{6457301876025}{2048}e^{24}b^{10}c^{9} - \\
-\frac{260186689035}{128}e^{26}b^{12}c^{7} + \frac{10886715}{8}e^{32}b^{18}c - \frac{72747675}{4096}c^{19}e^{14} )l_{15}(\psi) 
+(-\frac{13125105}{4}e^{32}b^{17}c^{2} + \frac{910922985}{16}c^{4}e^{30}b^{15} - \\
-\frac{537794462205}{8192}c^{14}e^{20}b^{5} 
+\frac{624776873175}{131072}c^{16}e^{18}b^{3} - \frac{734504175405}{1024}e^{24}b^{9}c^{10} 
+\frac{1355478329205}{4096}e^{22}b^{7}c^{12} +\\
+\frac{181153614915}{256}e^{26}b^{11}c^{8} 
-\frac{10038670425}{32}e^{28}b^{13}c^{6} + \frac{101745}{4}e^{34}b^{19} - \frac{5397877485}{65536}e^{16}bc^{18} )l_{16}(\psi) + \\
+(-\frac{377055}{4}e^{34}b^{18}c + \frac{13125105}{4}c^{3}e^{32}b^{16} + \frac{3027625965}{32}e^{28}b^{12}c^{7} 
+\frac{27804161385}{8192}c^{15}e^{20}b^{4} - \frac{113966507655}{4096}e^{22}b^{6}c^{13} + \\
+\frac{94673824245}{1024}e^{24}b^{8}c^{11} - \frac{35311721445}{256}e^{26}b^{10}c^{9} 
-\frac{456936795}{16}e^{30}b^{14}c^{5} + \frac{43648605}{65536}c^{19}e^{16} - \\
-\frac{17011051785}{131072}e^{18}b^{2}c^{17} )l_{17}(\psi) + (-\frac{1805465025}{128}e^{28}b^{11}c^{8} 
+\frac{107544465}{16}c^{6}e^{30}b^{13} - \frac{5397877485}{65536}e^{20}b^{3}c^{16} + \\
+\frac{4757697945}{4096}e^{22}b^{5}c^{14} 
+\frac{3477338865}{256}e^{26}b^{9}c^{10} + \frac{182971425}{131072}e^{18}bc^{18} + \frac{377055}{4}c^{2}e^{3}b^{17} - \\
-\frac{1995}{2}e^{36}b^{19} - \frac{12352555215}{2048}e^{24}b^{7}c^{12} -\frac{10886715}{8}e^{32}b^{15}c^{4} )l_{18}(\psi) 
+(\frac{1995}{2}e^{36}b^{18}c - \frac{130945815}{256}e^{26}b^{8}c^{11} + \\
+\frac{1526175}{8}e^{32}b^{14}c^{5} 
+\frac{101846745}{128}e^{28}b^{10}c^{9} - \frac{101745}{4}e^{34}b^{16}c^{3} - \frac{9258795}{16}e^{30}b^{12}c^{7} - \\
-\frac{72747675}{4096}e^{22}b^{4}c^{15} + \frac{305540235}{2048}e^{24}b^{6}c^{13} - \frac{440895}{131072}c^{19}e^{18} 
+\frac{43648605}{65536}e^{20}b^{2}c^{17} )l_{19}(\psi))).
\end{gather*}

\[\textbf{Appendix (Item 4). Expression for the Function }K_{18;20}\textbf{:}\]
\allowdisplaybreaks
\begin{gather*}
K_{18;20}(\psi)={}\frac{81342706477725}{34359738368}\{16[( - \frac{9}{646}c^{17}e^{2}b^{2} + \frac{2}{6137}c^{19} 
+\frac{2145}{311296}c^{3}e^{16}b^{16} + \frac{3}{19}c^{15}e^{4}b^{4} + \frac{819}{608}c^{11}e^{8}b^{8} + \frac{1287}{2432}c^{7}e^{12}b^{12} - \\
-\frac{105}{152}c^{13}e^{6}b^{6} - \frac{3003}{2432}c^{9}e^{10}b^{10} - \frac{3861}{38912}c^{5}e^{14}b^{14} 
-\frac{143}{1245184}ce^{18}b^{18} )l_{0}(\psi) + (-\frac{42471}{2432}e^{12}b^{11}c^{8} + \frac{33033}{1216}e^{10}b^{9}c^{10} + \\
+\frac{99099}{19456}e^{14}b^{13}c^{6} + \frac{14157}{622592}e^{18}b^{17}c^{2} - \frac{23595}{38912}e^{16}b^{15}c^{4} 
+\frac{11}{323}e^{2}bc^{18} - \frac{143}{1245184}e^{20}b^{19} - \frac{33}{38}e^{4}b^{3}c^{16} - \frac{3003}{152}e^{8}b^{7}c^{12} + \\
+\frac{495}{76}e^{6}b^{5}c^{14} )l_{1}(\psi) + (\frac{477}{323}c^{17}e^{4}b^{2} - \frac{3303}{152}e^{6}b^{4}c^{15} - \frac{9}{646}c^{19}e^{2}
+\frac{16443}{152}c^{13}e^{8}b^{6} - \frac{554463}{2432}e^{10}b^{8}c^{11} + \frac{16731}{76}e^{12}b^{10}c^{9} - \\
-\frac{3810807}{38912}e^{14}b^{12}c^{7} + \frac{14157}{622592}e^{20}b^{18}c + \frac{734877}{38912}e^{16}b^{14}c^{5}
-\frac{1669239}{1245184}e^{18}b^{16}c^{3} )l_{2}(\psi) + (\frac{1221}{38}c^{16}e^{6}b^{3} - \frac{43461}{152}e^{8}b^{5}c^{14} - \\
-\frac{33}{38}e^{4}bc^{18} + \frac{291291}{304}c^{12}e^{10}b^{7} - \frac{3421275}{2432}e^{12}b^{9}c^{10}
+\frac{9216207}{9728}c^{8}e^{14}b^{11} - \frac{11132121}{38912}e^{16}b^{13}c^{6} + \frac{2722863}{77824}c^{4}e^{18}b^{15} - \\
-\frac{1669239}{1245184}e^{20}b^{17}c^{2} + \frac{2145}{311296}e^{22}b^{19} )l_{3}(\psi) + (\frac{119649}{304}c^{15}e^{8}b^{4} 
-\frac{2696883}{1216}c^{13}e^{10}b^{6} - \frac{3303}{152}e^{6}b^{2}c^{17} + \frac{3}{19}c^{19}e^{4} + \\
+\frac{12332385}{2432}c^{11}e^{12}b^{8} - \frac{201822621}{38912}e^{14}b^{10}c^{9} 
+\frac{187426239}{77824}c^{7}e^{16}b^{12} - \frac{149442579}{311296}e^{18}b^{14}c^{5} + \\
+\frac{2722863}{77824}c^{3}e^{20}b^{16} - \frac{23595}{38912}e^{22}b^{18}c)l_{4}(\psi) + (\frac{1779723}{608}e^{10}b^{5}c^{14}
-\frac{13149279}{1216}c^{12}e^{12}b^{7} - \frac{43461}{152}e^{8}b^{3}c^{16} + \frac{331287957}{19456}c^{10}e^{14}b^{9} + \\
+\frac{495}{76}e^{6}bc^{18} - \frac{470224755}{38912}e^{16}b^{11}c^{8} + \frac{591521931}{155648}e^{18}b^{13}c^{6}
-\frac{149442579}{311296}e^{20}b^{15}c^{4} - \frac{3861}{38912}e^{24}b^{19} + \frac{734877}{38912}e^{22}b^{17}c^{2} )l_{5}(\psi) + \\
+(\frac{2112999}{152}c^{13}e^{12}b^{6} - \frac{1342122327}{38912}e^{14}b^{8}c^{11} - \frac{2696883}{1216}c^{15}e^{10}b^{4}
+\frac{1461815355}{38912}c^{9}e^{16}b^{10} + \frac{16443}{152}c^{17}e^{8}b^{2} - \\
-\frac{300248949}{16384}e^{18}b^{12}c^{7}
-\frac{105}{152}c^{19}e^{6} + \frac{591521931}{155648}e^{20}b^{14}c^{5} - \frac{11132121}{38912}e^{22}b^{16}c^{3} 
+\frac{99099}{19456}e^{24}b^{18}c)l_{6}(\psi) + \\
+(\frac{212110041}{4864}c^{12}e^{14}b^{7} - \frac{2872205193}{38912}e^{16}b^{9}c^{10}
-\frac{13149279}{1216}c^{14}e^{12}b^{5} + \frac{4322939049}{77824}c^{8}e^{18}b^{11} + \frac{291291}{304}c^{16}e^{10}b^{3} - \\
-\frac{300248949}{16384}e^{20}b^{13}c^{6} - \frac{3003}{152}e^{8}bc^{18} + \frac{187426239}{77824}c^{4}e^{22}b^{15}
-\frac{3810807}{38912}e^{24}b^{17}c^{2} + \frac{1287}{2432}e^{26}b^{19} )l_{7}(\psi) + (\frac{14382410211}{155648}c^{11}e^{16}b^{8} - \\
-\frac{66989234169}{622592}c^{9}e^{18}b^{10} - \frac{1342122327}{38912}e^{14}b^{6}c^{13}
+\frac{4322939049}{77824}c^{7}e^{20}b^{12} + \frac{12332385}{2432}c^{15}e^{12}b^{4} - \\
-\frac{470224755}{38912}e^{22}b^{14}c^{5} - \frac{554463}{2432}e^{10}b^{2}c^{17} + \frac{819}{608}c^{19}e^{8}
+\frac{9216207}{9728}c^{3}e^{24}b^{16} - \frac{42471}{2432}e^{26}b^{18}c)l_{8}(\psi) + (\frac{2196652029}{16384}c^{10}e^{18}b^{9} - \\
-\frac{66989234169}{622592}c^{8}e^{20}b^{11} - \frac{2872205193}{38912}e^{16}b^{7}c^{12} 
+\frac{1461815355}{38912}c^{6}e^{22}b^{13} + \frac{331287957}{19456}c^{14}e^{14}b^{5} - \\
-\frac{201822621}{38912}e^{24}b^{15}c^{4} - \frac{3421275}{2432}e^{12}b^{3}c^{16} + \frac{16731}{76}e^{26}b^{17}c^{2} 
+\frac{33033}{1216}e^{10}bc^{18} - \frac{3003}{2432}e^{28}b^{19} )l_{9}(\psi) + (\frac{2196652029}{16384}c^{9}e^{20}b^{10} - \\
-\frac{2872205193}{38912}e^{22}b^{12}c^{7} - \frac{66989234169}{622592}c^{11}e^{18}b^{8}
+\frac{331287957}{19456}c^{5}e^{24}b^{14} + \frac{1461815355}{38912}c^{13}e^{16}b^{6} - \\
-\frac{3421275}{2432}e^{26}b^{16}c^{3} - \frac{201822621}{38912}e^{14}b^{4}c^{15} + \frac{33033}{1216}e^{28}b^{18}c 
+\frac{16731}{76}e^{12}b^{2}c^{17} - \frac{3003}{2432}c^{19}e^{10} )l_{10}(\psi) + (\frac{14382410211}{155648}c^{8}e^{22}b^{11} - \\
-\frac{1342122327}{38912}e^{24}b^{13}c^{6} - \frac{66989234169}{622592}c^{10}e^{20}b^{9} 
+\frac{12332385}{2432}c^{4}e^{26}b^{15} + \frac{4322939049}{77824}c^{12}e^{18}b^{7} - \frac{554463}{2432}e^{28}b^{17}c^{2} - \\
-\frac{470224755}{38912}e^{16}b^{5}c^{14} + \frac{819}{608}e^{30}b^{19} + \frac{9216207}{9728}c^{16}e^{14}b^{3}
-\frac{42471}{2432}e^{12}bc^{18})l_{11}(\psi) + (\frac{212110041}{4864}c^{7}e^{24}b^{12} -\frac{13149279}{1216}c^{5}e^{26}b^{14} - \\
-\frac{2872205193}{38912}e^{22}b^{10}c^{9} + \frac{291291}{304}c^{3}e^{28}b^{16} + \frac{4322939049}{77824}c^{11}e^{20}b^{8}
-\frac{3003}{152}e^{30}b^{18}c - \frac{300248949}{16384}e^{18}b^{6}c^{13} + \frac{187426239}{77824}c^{15}e^{16}b^{4} - \\
-\frac{3810807}{38912}e^{14}b^{2}c^{17} + \frac{1287}{2432}c^{19}e^{12} )l_{12}(\psi) + (\frac{2112999}{152}c^{6}e^{26}b^{13}
-\frac{2696883}{1216}c^{4}e^{28}b^{15} -\frac{1342122327}{38912}e^{24}b^{11}c^{8} + \frac{16443}{152}c^{2}e^{30}b^{17} + \\
+\frac{1461815355}{38912}c^{10}e^{22}b^{9} - \frac{300248949}{16384}e^{20}b^{7}c^{12} - \frac{105}{152}e^{32}b^{19}
+\frac{591521931}{155648}e^{18}b^{5}c^{14} - \frac{11132121}{38912}e^{16}b^{3}c^{16} + \\
+\frac{99099}{19456}e^{14}bc^{18})l_{13}(\psi) + (\frac{1779723}{608}e^{28}b^{14}c^{5} - \frac{43461}{152}e^{30}b^{16}c^{3}
-\frac{13149279}{1216}c^{7}e^{26}b^{12} + \frac{495}{76}e^{32}b^{18}c + \frac{331287957}{19456}c^{9}e^{24}b^{10} -\\
-\frac{470224755}{38912}e^{22}b^{8}c^{11} + \frac{591521931}{155648}e^{20}b^{6}c^{13}
-\frac{149442579}{311296}e^{18}b^{4}c^{15} - \frac{3861}{38912}c^{19}e^{14} + \frac{734877}{38912}e^{16}b^{2}c^{17} )l_{14}(\psi) + \\
+(\frac{119649}{304}c^{4}e^{30}b^{15} - \frac{3303}{152}e^{32}b^{17}c^{2} - \frac{2696883}{1216}c^{6}e^{28}b^{13}
+\frac{3}{19}e^{34}b^{19} + \frac{12332385}{2432}c^{8}e^{26}b^{11} - \frac{201822621}{38912}e^{24}b^{9}c^{10} + \\
+\frac{187426239}{77824}c^{12}e^{22}b^{7} - \frac{149442579}{311296}e^{20}b^{5}c^{14}
+\frac{2722863}{77824}c^{16}e^{18}b^{3} - \frac{23595}{38912}e^{16}bc^{18} )l_{15}(\psi) + (\frac{1221}{38}c^{3}e^{32}b^{16} - \\
-\frac{33}{38}e^{34}b^{18}c -\frac{43461}{152}e^{30}b^{14}c^{5} + \frac{291291}{304}c^{7}e^{28}b^{12}
-\frac{3421275}{2432}e^{26}b^{10}c^{9} + \frac{9216207}{9728}c^{11}e^{24}b^{8} -\frac{11132121}{38912}e^{22}b^{6}c^{13} + \\
+\frac{2722863}{77824}c^{15}e^{20}b^{4} - \frac{1669239}{1245184}e^{18}b^{2}c^{17}
+\frac{2145}{311296}c^{19}e^{16} )l_{16}(\psi) + (\frac{477}{323}c^{2}e^{34}b^{17} - \frac{3303}{152}e^{32}b^{15}c^{4} - \\
-\frac{9}{646}e^{36}b^{19} + \frac{16443}{152}c^{6}e^{30}b^{13} - \frac{554463}{2432}e^{28}b^{11}c^{8} + \frac{16731}{76}e^{26}b^{9}c^{10}
-\frac{3810807}{38912}e^{24}b^{7}c^{12} + \frac{734877}{38912}e^{22}b^{5}c^{14} -\frac{1669239}{1245184}e^{20}b^{3}c^{16} + \\
+\frac{14157}{622592}e^{18}bc^{18} )l_{17}(\psi) + (\frac{33033}{1216}e^{28}b^{10}c^{9} - \frac{42471}{2432}e^{26}b^{8}c^{11}
+\frac{99099}{19456}e^{24}b^{6}c^{13} - \frac{23595}{38912}e^{22}b^{4}c^{15} + \frac{14157}{622592}e^{20}b^{2}c^{17} + \\
+\frac{11}{323}e^{36}b^{18}c - \frac{33}{38}e^{34}b^{16}c^{3} + \frac{495}{76}e^{32}b^{14}c^{5} -\frac{143}{1245184}c^{19}e^{18}
-\frac{3003}{152}e^{30}b^{12}c^{7} )l_{18}(\psi) + (\frac{2}{6137}e^{38}b^{19} + \frac{3}{19}e^{34}b^{15}c^{4} + \\
+\frac{819}{608}e^{30}b^{11}c^{8}
+\frac{1287}{2432}e^{26}b^{7}c^{12} - \frac{105}{152}e^{32}b^{13}c^{6} - \frac{3003}{2432}e^{28}b^{9}c^{10}
-\frac{3861}{38912}e^{24}b^{5}c^{14} - \frac{9}{646}e^{36}b^{17}c^{2} -\frac{143}{1245184}e^{20}bc^{18} + \\
+\frac{2145}{311296}e^{22}b^{3}c^{16} ) l_{19}(\psi)] + i\Omega e60[( \frac{3}{76}e^{2}b^{3}c^{16} -\frac{1}{646}bc^{18}
-\frac{1287}{1245184}e^{16}b^{17}c^{2} - \frac{45}{152}e^{4}b^{5}c^{14} - \frac{3003}{2432}e^{8}b^{9}c^{10} - \\
-\frac{9009}{38912}e^{12}b^{13}c^{6} + \frac{273}{304}e^{6}b^{7}c^{12} + \frac{3861}{4864}e^{10}b^{11}c^{8}
+\frac{2145}{77824}e^{14}b^{15}c^{4} + \frac{13}{2490368}e^{18}b^{19} )l_{0}(\psi) + (-\frac{102531}{4864}c^{9}e^{10}b^{10} + \\
+\frac{364221}{38912}c^{7}e^{12}b^{12} - \frac{140283}{77824}c^{5}e^{14}b^{14} + \frac{159159}{1245184}c^{3}e^{16}b^{16}
-\frac{5395}{2490368}ce^{18}b^{18} + \frac{1}{646}c^{19} - \frac{189}{1292}c^{17}e^{2}b^{2} + \frac{321}{152}c^{15}e^{4}b^{4} - \\
-\frac{3171}{304}c^{13}e^{6}b^{6} + \frac{53235}{2432}c^{11}e^{8}b^{8})l_{1}(\psi) + (\frac{189}{1292}e^{2}bc^{18} - \frac{387}{76}e^{4}b^{3}c^{16}
+\frac{13473}{304}e^{6}b^{5}c^{14} - \frac{89271}{608}e^{8}b^{7}c^{12} + \frac{1041183}{4864}e^{10}b^{9}c^{10} - \\
-\frac{2791503}{19456}e^{12}b^{11}c^{8} + \frac{3360357}{77824}e^{14}b^{13}c^{6} - \frac{819819}{155648}e^{16}b^{15}c^{4}
+\frac{501579}{2490368}e^{18}b^{17}c^{2} - \frac{1287}{1245184}e^{20}b^{19} )l_{2}(\psi) + (-\frac{26859}{304}e^{6}b^{4}c^{15} + \\
+\frac{387}{76}c^{17}e^{4}b^{2} - \frac{3}{76}c^{19}e^{2} + \frac{295995}{608}c^{13}e^{8}b^{6} - \frac{5341401}{4864}e^{10}b^{8}c^{11}
+\frac{1139853}{1024}e^{12}b^{10}c^{9} - \frac{39966927}{77824}e^{14}b^{12}c^{7} + \frac{15857127}{155648}e^{16}b^{14}c^{5} - \\
-\frac{18421845}{2490368}e^{18}b^{16}c^{3} + \frac{159159}{1245184}e^{20}b^{18}c)l_{3}(\psi)
+(-\frac{1061829}{1216}e^{8}b^{5}c^{14} + \frac{26859}{304}c^{16}e^{6}b^{3} + \frac{7671729}{2432}c^{12}e^{10}b^{7} - \\
-\frac{321}{152}e^{4}bc^{18} - \frac{190396635}{38912}e^{12}b^{9}c^{10} + \frac{267360093}{77824}c^{8}e^{14}b^{11}
-\frac{333642309}{311296}e^{16}b^{13}c^{6} + \frac{83770557}{622592}c^{4}e^{18}b^{15} - \\
-\frac{819819}{155648}e^{20}b^{17}c^{2} + \frac{2145}{77824}e^{22}b^{19} )l_{4}(\psi) + (\frac{1061829}{1216}c^{15}e^{8}b^{4}
-\frac{12897927}{2432}c^{13}e^{10}b^{6} + \frac{45}{152}c^{19}e^{4} - \frac{13473}{304}e^{6}b^{2}c^{17} + \\
+\frac{26345709}{2048}c^{11}e^{12}b^{8} - \frac{1072566495}{77824}e^{14}b^{10}c^{9}
+\frac{2066997933}{311296}c^{7}e^{16}b^{12} - \frac{849081987}{622592}e^{18}b^{14}c^{5} + \\
+\frac{15857127}{155648}c^{3}e^{20}b^{16} - \frac{140283}{77824}e^{22}b^{18}c)l_{5}(\psi) 
+(-\frac{201920901}{9728}c^{12}e^{12}b^{7} + \frac{12897927}{2432}e^{10}b^{5}c^{14} - \frac{295995}{608}e^{8}b^{3}c^{16} + \\
+\frac{2671045377}{77824}c^{10}e^{14}b^{9} - \frac{3947896953}{155648}e^{16}b^{11}c^{8} + \frac{3171}{304}e^{6}bc^{18}
+\frac{5135974935}{622592}e^{18}b^{13}c^{6} - \frac{333642309}{311296}e^{20}b^{15}c^{4} + \\
+\frac{3360357}{77824}e^{22}b^{17}c^{2} - \frac{9009}{38912}e^{24}b^{19} )l_{6}(\psi)
+(-\frac{4201128243}{77824}e^{14}b^{8}c^{11} + \frac{201920901}{9728}c^{13}e^{12}b^{6} - \frac{273}{304}c^{19}e^{6} + \\
+\frac{9545130309}{155648}c^{9}e^{16}b^{10} - \frac{7671729}{2432}c^{15}e^{10}b^{4}
-\frac{19317776361}{622592}e^{18}b^{12}c^{7} + \frac{89271}{608}c^{17}e^{8}b^{2}
+\frac{2066997933}{311296}e^{20}b^{14}c^{5} -\\
-\frac{39966927}{77824}e^{22}b^{16}c^{3}
+\frac{364221}{38912}e^{24}b^{18}c)l_{7}(\psi) + (-\frac{59214472317}{622592}e^{16}b^{9}c^{10}
+\frac{4201128243}{77824}c^{12}e^{14}b^{7} + \frac{92536841475}{1245184}c^{8}e^{18}b^{11} - \\
-\frac{26345709}{2048}c^{14}e^{12}b^{5} - \frac{3947896953}{155648}e^{20}b^{13}c^{6} + \frac{5341401}{4864}c^{16}e^{10}b^{3}
-\frac{53235}{2432}e^{8}bc^{18} + \frac{267360093}{77824}c^{4}e^{22}b^{15} - \frac{2791503}{19456}e^{24}b^{17}c^{2} + \\
+\frac{3861}{4864}e^{26}b^{19} )l_{8}(\psi) + (-\frac{142873576989}{1245184}c^{9}e^{18}b^{10}
+\frac{59214472317}{622592}c^{11}e^{16}b^{8} + \frac{3003}{2432}c^{19}e^{8} + \frac{9545130309}{155648}c^{7}e^{20}b^{12} - \\
-\frac{2671045377}{77824}e^{14}b^{6}c^{13} - \frac{1072566495}{77824}e^{22}b^{14}c^{5}
+\frac{190396635}{38912}c^{15}e^{12}b^{4} - \frac{1041183}{4864}e^{10}b^{2}c^{17} + \frac{1139853}{1024}c^{3}e^{24}b^{16} - \\
-\frac{102531}{4864}e^{26}b^{18}c)l_{9}(\psi) + (-\frac{59214472317}{622592}c^{8}e^{20}b^{11}
+\frac{142873576989}{1245184}c^{10}e^{18}b^{9} - \frac{3003}{2432}e^{28}b^{19} + \\
+\frac{2671045377}{77824}c^{6}e^{22}b^{13} - \frac{9545130309}{155648}e^{16}b^{7}c^{12}
-\frac{190396635}{38912}e^{24}b^{15}c^{4} + \frac{1072566495}{77824}c^{14}e^{14}b^{5} - \frac{1139853}{1024}e^{12}b^{3}c^{16} + \\
+\frac{1041183}{4864}e^{26}b^{17}c^{2} + \frac{102531}{4864}e^{10}bc^{18} )l_{10}(\psi)
+(-\frac{4201128243}{77824}e^{22}b^{12}c^{7} + \frac{59214472317}{622592}c^{9}e^{20}b^{10} + \\
+\frac{26345709}{2048}c^{5}e^{24}b^{14} - \frac{92536841475}{1245184}c^{11}e^{18}b^{8} - \frac{5341401}{4864}e^{26}b^{16}c^{3}
+\frac{3947896953}{155648}c^{13}e^{16}b^{6} - \frac{267360093}{77824}e^{14}b^{4}c^{15} +\\
+\frac{53235}{2432}e^{28}b^{18}c
+\frac{2791503}{19456}e^{12}b^{2}c^{17} - \frac{3861}{4864}c^{19}e^{10} )l_{11}(\psi) + (-\frac{201920901}{9728}e^{24}b^{13}c^{6} + \\
+\frac{4201128243}{77824}c^{8}e^{22}b^{11} + \frac{273}{304}e^{30}b^{19} + \frac{7671729}{2432}c^{4}e^{26}b^{15}
-\frac{9545130309}{155648}c^{10}e^{20}b^{9} - \frac{89271}{608}e^{28}b^{17}c^{2} + \frac{19317776361}{622592}c^{12}e^{18}b^{7} - \\
-\frac{2066997933}{311296}e^{16}b^{5}c^{14} + \frac{39966927}{77824}c^{16}e^{14}b^{3} - \frac{364221}{38912}e^{12}bc^{18} )l_{12}(\psi)
+(-\frac{12897927}{2432}c^{5}e^{26}b^{14} + \frac{201920901}{9728}c^{7}e^{24}b^{12} +\\
+\frac{333642309}{311296}c^{15}e^{16}b^{4}
-\frac{2671045377}{77824}e^{22}b^{10}c^{9} + \frac{295995}{608}c^{3}e^{28}b^{16} - \frac{3171}{304}e^{30}b^{18}c
+\frac{3947896953}{155648}c^{11}e^{20}b^{8} -\\
-\frac{5135974935}{622592}e^{18}b^{6}c^{13} - \frac{3360357}{77824}e^{14}b^{2}c^{17}
+\frac{9009}{38912}c^{19}e^{12} )l_{13}(\psi) + (\frac{12897927}{2432}c^{6}e^{26}b^{13} - \frac{1061829}{1216}c^{4}e^{28}b^{15} - \\
-\frac{45}{152}e^{32}b^{19} - \frac{26345709}{2048}e^{24}b^{11}c^{8} + \frac{13473}{304}c^{2}e^{30}b^{17}
+\frac{1072566495}{77824}c^{10}e^{22}b^{9} - \frac{2066997933}{311296}e^{20}b^{7}c^{12} + \\
+\frac{849081987}{622592}e^{18}b^{5}c^{14} - \frac{15857127}{155648}e^{16}b^{3}c^{16} + \frac{140283}{77824}e^{14}bc^{18} )l_{14}(\psi)
+(-\frac{26859}{304}e^{30}b^{16}c^{3} + \frac{1061829}{1216}e^{28}b^{14}c^{5} + \frac{321}{152}e^{32}b^{18}c - \\
-\frac{7671729}{2432}c^{7}e^{26}b^{12} + \frac{190396635}{38912}c^{9}e^{24}b^{10} - \frac{267360093}{77824}e^{22}b^{8}c^{11}
+\frac{333642309}{311296}e^{20}b^{6}c^{13} - \frac{83770557}{622592}e^{18}b^{4}c^{15} + \frac{819819}{155648}e^{16}b^{2}c^{17} - \\
-\frac{2145}{77824}c^{19}e^{14} )l_{15}(\psi) + (-\frac{387}{76}e^{32}b^{17}c^{2} + \frac{26859}{304}c^{4}e^{30}b^{15} + \frac{3}{76}e^{34}b^{19}
-\frac{295995}{608}c^{6}e^{28}b^{13} + \frac{5341401}{4864}c^{8}e^{26}b^{11} - \\
-\frac{1139853}{1024}e^{24}b^{9}c^{10}
+\frac{39966927}{77824}c^{12}e^{22}b^{7} - \frac{15857127}{155648}e^{20}b^{5}c^{14} + \frac{18421845}{2490368}c^{16}e^{18}b^{3}
-\frac{159159}{1245184}e^{16}bc^{18} )l_{16}(\psi) +\\
+(-\frac{189}{1292}e^{34}b^{18}c + \frac{387}{76}c^{3}e^{32}b^{16}
-\frac{13473}{304}e^{30}b^{14}c^{5} + \frac{89271}{608}c^{7}e^{28}b^{12} - \frac{1041183}{4864}e^{26}b^{10}c^{9} + \\
+\frac{2791503}{19456}c^{11}e^{24}b^{8} - \frac{3360357}{77824}e^{22}b^{6}c^{13} + \frac{819819}{155648}c^{15}e^{20}b^{4}
-\frac{501579}{2490368}e^{18}b^{2}c^{17} + \frac{1287}{1245184}c^{19}e^{16} )l_{17}(\psi) + (-\frac{364221}{38912}e^{24}b^{7}c^{12} + \\
+\frac{102531}{4864}e^{26}b^{9}c^{10} - \frac{159159}{1245184}e^{20}b^{3}c^{16} + \frac{140283}{77824}e^{22}b^{5}c^{14}
+\frac{5395}{2490368}e^{18}bc^{18} + \frac{189}{1292}c^{2}e^{34}b^{17} - \frac{1}{646}e^{36}b^{19} + \frac{3171}{304}c^{6}e^{30}b^{13} - \\
-\frac{321}{152}e^{32}b^{15}c^{4} - \frac{53235}{2432}e^{28}b^{11}c^{8} )l_{18}(\psi) + (\frac{1}{646}e^{36}b^{18}c + \frac{45}{152}e^{32}b^{14}c^{5}
+\frac{3003}{2432}e^{28}b^{10}c^{9} + \frac{9009}{38912}e^{24}b^{6}c^{13} - \frac{273}{304}e^{30}b^{12}c^{7} -\\
-\frac{3861}{4864}e^{26}b^{8}c^{11}
-\frac{2145}{77824}e^{22}b^{4}c^{15} - \frac{3}{76}e^{34}b^{16}c^{3} - \frac{13}{2490368}c^{19}e^{18} + \frac{1287}{1245184}e^{20}b^{2}c^{17} ) l_{19}(\psi)] \}.
\end{gather*}
Where\\

$e=\sqrt{{\rm 1}-{\rm \lambda}^{{2}}},\qquad b=sin\psi,\qquad c=\lambda+cos\psi$.
\noindent \textbf{}
\noindent 
\pagebreak
\end{document}